\documentclass[aps,prb,reprint,amsmath,amssymb,showpacs,superscriptaddress,twocolumn]{revtex4}
\usepackage{graphicx}
\usepackage{xcolor}
\usepackage{subfigure}
\usepackage{hyperref,hypcap}
\usepackage{braket}
\usepackage{amsmath}

\begin{document}
\title{Theory of Two-Dimensional Spatially Indirect Equilibrium Exciton Condensates}
\author{Feng-Cheng Wu$^{\dagger}$}
\author{Fei Xue$^{\dagger}$}
\author{A.H. MacDonald}
\affiliation{Department of Physics, The University of Texas at Austin, Austin, TX 78712, USA\\
\rm{($^{\dagger}$These authors contributed equally to this work.)}}

\date{\today}

\begin{abstract}
	We present a theory of bilayer two-dimensional electron systems that host a
	spatially indirect exciton condensate when in thermal equilibrium.  
	Equilibrium bilayer exciton condensates (BXCs)
	are expected to form when two nearby semiconductor layers are electrically 
	isolated, and when the 
	conduction band of one layer is brought close to degeneracy with the valence band of a nearby layer by varying bias or gate voltages.
	BXCs are characterized by spontaneous inter-layer phase coherence
	and counterflow superfluidity. The bilayer system we consider is composed of 
	two transition metal dichalcogenide monolayers separated and surrounded 
	by hexagonal boron nitride. We use mean-field-theory and a bosonic weakly interacting 
	exciton model to explore the BXC phase diagram, and time-dependent mean-field 
	theory to address condensate collective mode spectra and quantum fluctuations.
	We find that a phase transition
	occurs between states containing one and two condensate components as the layer separation and 
	the exciton density are varied, and derive simple approximate expressions for the 
	exciton-exciton interaction strength which we show can be measured capacitively.
\end{abstract}

\pacs{71.35.-y, 73.21.-b}

\maketitle

\section{Introduction}
Recent advances in the study of two-dimensional van der Waals materials\cite{Geim2013} have opened up
new horizons in condensed matter physics by allowing familiar properties, including those of metals, superconductors, gapless
semiconductors, semiconductors, and insulators, to be combined in new ways simply by designing stacks
of atomically thick layers.  In this article we consider condensation of spatially indirect excitons
in the case of a two-dimensional semiconductor bilayer formed by 
two group-VI transition metal dichalcogenides (TMD) that are separated and surrounded by an insulator,
for example hexagonal boron nitride (hBN). The TMDs are in their 2H structure monolayer form.
Two-dimensional material stacks of this type are     
promising hosts for exciton condensation, both because they host strongly bound excitons,
\cite{Reichman, Steven13, PL_Rana, Heinz_14, Dark_WS2, Dark_WSe2, Wu2015}
and because of recent progress in realizing flexible high quality TMD 
heterostructures.\cite{Fang2014,FWang2014, Duan2014, Gong2014, Cao2015, Rivera2015}

\begin{figure}[t]
	\includegraphics[width=0.9\columnwidth]{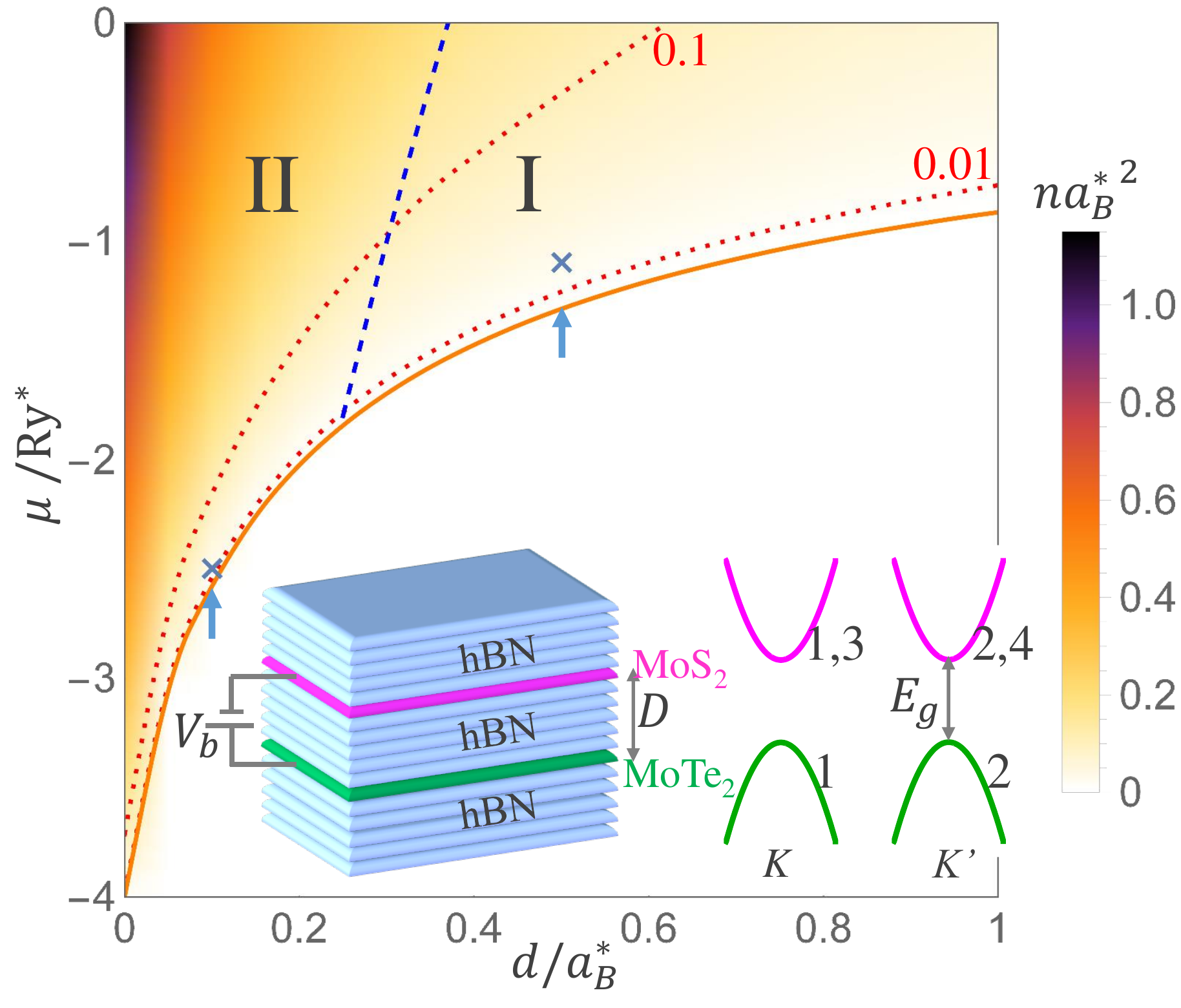}
	\caption{(Color online) Transferred charge density and exciton condensate phase as a function of
		effective layer separation $d$ and chemical potential $\mu$. The solid orange line marks the second-order phase transition
		from the phase in which there is no charge transfer between the bilayers
		{\it i.e.} the phase with no excitons present, to the bilayer exciton condensate (BXC) phase. 
		The blue dashed line separates phase-II in which two condensate flavors are present  
		from phase-I in which the ground state has a single condensate flavor.
		The two red dotted lines are contours at the density values $n a_B^{*2}=$ 0.01 and 0.1.
		(For $ n a_B^{*2} \gtrsim 0.1$ the 
		exciton-condensate ground state is expected to be superseded by an electron-hole plasma state.
		See text for a more complete discussion.)  
		The arrows mark the values of $d$ examined in Fig.~\ref{Fig:Energy}(a) and \ref{Fig:Energy}(b), and the
		crosses($\times$) indicate the parameter 
		values examined in Fig.~\ref{Fig:Energy}(c) and \ref{Fig:Energy}(d).
		The left inset is a schematic experimental setup for BXC studies in which the spatially indirect gap 
		is tuned by an interlayer bias potential, and the right inset illustrates 
		the bilayer band structure in the absence of the bias potential $V_b$.  }
	\label{Fig:PhaseDiagram}
\end{figure}

In van der Waals heterostructures it is possible\cite{Tutuc2014,Tutuc2015} to tune the positions of the
Fermi levels in individual layers over wide ranges while maintaining overall charge neutrality, either by
applying a gate voltage between surrounding electrodes or a bias voltage between the semiconductor layers.
When the indirect band gap between the conduction band of one layer and the valence
band of the other layer is reduced to less than the indirect exciton binding energy, 
charge will be transferred between layers in equilibrium.
At low densities, the transferred charges form spatially indirect
excitons, and these are 
expected\cite{Lozovik1976,Zhu1995,Lozovik1996,Fernandez1996,Vina1999,Combescot2008a,
Combescot2008b,Butov2012} to form Bose condensates.  
The bilayer exciton condensate (BXC) state has spontaneous interlayer phase coherence and 
supports dissipationless counterflow supercurrents\cite{Eisenstein2004,Su2008} 
that could enable the design of low-dissipation electronic devices.\cite{Banerjee2009}

Exciton condensates in TMD heterostructures are similar to atomic
spinor Bose-Einstein condensates because of the presence of both spin and valley degrees of freedom. 
The spin-valley coupling of conduction band electrons and valence band holes
that are specific to TMD heterostructures\cite{Yao12} enriches the excitonic physics.
In this paper we study the interplay between the exciton condensation
and spin and valley internal degrees of freedom to construct an exciton condensate zero temperature phase diagram
as a function of effective layer separation $d$ and exciton chemical potential $\mu$, or equivalently exciton density.
We demonstrate that there are two distinct condensate phases with different number of condensate flavors, as shown in Fig.~\ref{Fig:PhaseDiagram}.

Our paper is organized as follows. In Sec.~\ref{Meanfield}, we explain how we model the heterostructure, and present the mean-field phase diagram implied by Hartree-Fock theory. In Sec.~\ref{BoseM}, we derive an effective boson model that  
incorporates exciton-exciton interaction effects and 
can be used to describe excitons in the low density limit. 
The difference between the strengths of the repulsive interactions between excitons with the same internal label and
between excitons with different internal labels changes sign as the layer separation increases.  
This change drives the transition from phase-II, a phase with two condensate flavors present, to phase-I, 
a phase with only one condensate flavor. 
Both phases spontaneously break the symmetry of the model Hamiltonian, and the symmetry breaking pattern of each phase is analyzed. 
In this section we also explain how capacitance measurements can be used to study the exciton phase diagram experimentally and 
to extract the value of the exciton-exciton interaction strength within each phase. 
In Sec.~\ref{Fluc}, we use a time-dependent Hartree-Fock theory to study the stability of phase-I against 
small fluctuations, and to calculate the collective mode spectra of these exciton condensates.
Finally in Sec.~\ref{summ}, we present a brief summary, discuss issues related to experiments, and comment 
on the relationship between our work and previous studies.

\section{Mean-Field Phase Diagram}
\label{Meanfield}
We consider two monolayer TMD semiconductors separated and surrounded by hBN (Fig.~\ref{Fig:PhaseDiagram}).
Many of the points we make apply with minor modification, however, to 
any bilayer two-dimensional semiconductor system.  
Monolayer TMDs are direct-gap semiconductors with band extrema located at valleys $K$ and $K'$.
Because these TMD layers lack inversion symmetry, spin degeneracy in the TMD bands is lifted by spin-orbit interactions.   
Because of differences between the orbital character of conduction and valence band states,\cite{Yao12} it turns out that 
spin splitting is large at the valence band maxima and small at the conduction band minima.
As illustrated in Fig.~\ref{Fig:PhaseDiagram}, we therefore retain in our theory the two valley-degenerate valence bands
labeled by $v=1,2$, and four conduction bands with labels $c=1,2,3,4$ corresponding to spin and valley.  
We assume that exciton binding energies and densities are small enough to justify a 
parabolic band approximation for all band extrema.  
Our mean-field {\it ansatz} allows up to two types of excitons to be present;
for examples pairs formed from holes in band $v=1$ and electrons, selected by spin-splitting, in band $c=1$, 
can condense, along with pairs formed from holes in band $v=2$ and electrons in band $c=2$.  
Although the unpaired conduction bands $c=3,4$ are only slightly higher in energy,
this pairing {\it ansatz} is fully self-consistent at low exciton density, because of the substantial exciton binding energy.
Our pairing {\it ansatz} is also justified by an interacting boson model, described in Sec.~\ref{BoseM}, which allows 
for the most general possible pairing scenario.  

The {\it ansatz} leads to the mean-field Hamiltonian:  
\begin{equation}
\begin{aligned}
H_{MF}=&\sum_{\vec{k}(v c)}'(a_{c\vec{k}}^{\dagger}, a_{v\vec{k}}^\dagger)
(\zeta_{\vec{k}}+\xi_{\vec{k}}^{(vc)} \sigma_z-\Delta_{\vec{k}}^{(vc)} \sigma_x)
\begin{pmatrix} a_{c\vec{k}}\\a_{v\vec{k}}\end{pmatrix}\\
+&\sum_{\vec{k},c=3,4}
\big(\frac{\hbar^2k^2}{2m_e}-\frac{1}{2}\tilde{\mu}\big)
a_{c\vec{k}}^{\dagger}a_{c\vec{k}},
\end{aligned}
\label{MF}
\end{equation}
where the prime in the first summation restricts the pair index $(vc)$ to (11) and (22) contributions.
$a^\dagger$ and $a$ are fermionic creation and annihilation operators.
The kinetic term $\zeta_{\vec{k}}=\hbar^2k^2(1/(4m_e)-1/(4m_h))$
accounts for the difference between conduction and valence band effective masses,
$m_e$ and $m_h$, and $\sigma_{z,x}$ are Pauli matrices. 
The dressed energy difference between conduction and valence bands, $\xi_{\vec{k}}^{(vc)}$, and 
the coherence induced effective interlayer tunneling amplitude, $\Delta_{\vec{k}}^{(vc)}$, are defined as:
\begin{equation}
\begin{aligned}
\xi_{\vec{k}}^{(vc)}=&\frac{\hbar^2k^2}{4m}-\frac{1}{2}\tilde{\mu}
-\frac{1}{A}\sum_{\vec{k}'}V(\vec{k}-\vec{k}')\langle a_{c\vec{k}'}^{\dagger}a_{c\vec{k}'} \rangle,\\
\Delta_{\vec{k}}^{(vc)}=&\frac{1}{A}\sum_{\vec{k}'}U(\vec{k}-\vec{k}')\langle a_{c \vec{k}'}^{\dagger}a_{v\vec{k}'} \rangle,
\end{aligned}
\label{SC}
\end{equation}
where $\langle...\rangle$ is the expectation value in the mean-field ground state,
\begin{equation}
\begin{aligned}
&\langle a_{c\vec{k}}^{\dagger}a_{c\vec{k}} \rangle=\frac{1}{2}(1-\xi_{\vec{k}}^{(vc)}/E_{\vec{k}}^{(vc)}),\\
&\langle a_{c\vec{k}}^{\dagger}a_{v\vec{k}} \rangle=\Delta_{\vec{k}}^{(vc)}/(2E_{\vec{k}}^{(vc)}),\\
&E_{\vec{k}}^{(vc)}=\sqrt{\xi_{\vec{k}}^{(vc)2}+\Delta_{\vec{k}}^{(vc)2}}.
\end{aligned}
\label{SC_2}
\end{equation}

In Eq. (\ref{SC}), $m=m_em_h/(m_e+m_h)$ is the reduced mass, and $A$ is the area of the system.
The paramter $\tilde{\mu}$ is:
\begin{equation}
\tilde{\mu}=\mu-4\pi e^2 n d /\epsilon, n=\frac{1}{A}\sum_{\vec{k}}\sum_{c=1,2} \langle a_{c \vec{k}}^\dagger a_{c \vec{k}} \rangle,
\label{SC_3}
\end{equation}
where $\mu$ is the chemical potential for excitons,
and $n$ is the total charge density transferred between layers.
Equations (\ref{SC}), (\ref{SC_2}) and (\ref{SC_3}) form a set of mean-field equations that can be solved self-consistently. 
Note that the (11) and (22) pairing channels are coupled through the dependence of $\tilde{\mu}$ on the total transferred density $n$.

The exciton chemical potential can be tuned electrically by applying a bias potential $V_b$ between
the electrically isolated layers: $\mu=V_b-E_{g}$
where $E_{g}$ is the spatially indirect band gap between the conduction band of the electron layer and 
the valence band of the hole layer. The band gap $E_g$ can be adjusted to a conveniently small value 
by choosing two-dimensional materials with favorable band alignments\cite{Cho2013, chiu2015}.

$V(\vec{q})=2\pi e^2/(\epsilon q)$ and $U(\vec{q})=V(\vec{q})e^{-qd}$
are the Coulomb interaction potentials within and between layers. The forms of
Coulomb potentials are determined by solving the Poisson equation for our schematic experimental 
setup(Fig.~\ref{Fig:PhaseDiagram}). $\epsilon=\sqrt{\epsilon_{\perp}\epsilon_{\parallel}}$, where $\epsilon_{\perp}$ and $\epsilon_{\parallel}$
are hBN dielectric constants perpendicular and parallel to the z-axis, is the effective dielectric constant
due to insulator layer(hBN) between electron and hole layers. $d=D\sqrt{\epsilon_{\perp}/\epsilon_{\parallel}}$, where $D$ is the geometric
layer separation between electron and hole layers, is the effective layer separation and slightly larger than $D$.\cite{Cai2007}

Below we express lengths and energies in terms of the characteristic scales
$a_B^*=\epsilon\hbar^2/(m e^2)$, and $\text{Ry}^*=e^2/(2\epsilon a_B^*)$. 
Typical values for different material combinations are listed in Table~\ref{Parameters}.

\begin{figure}[t]
	\includegraphics[width=0.9\columnwidth]{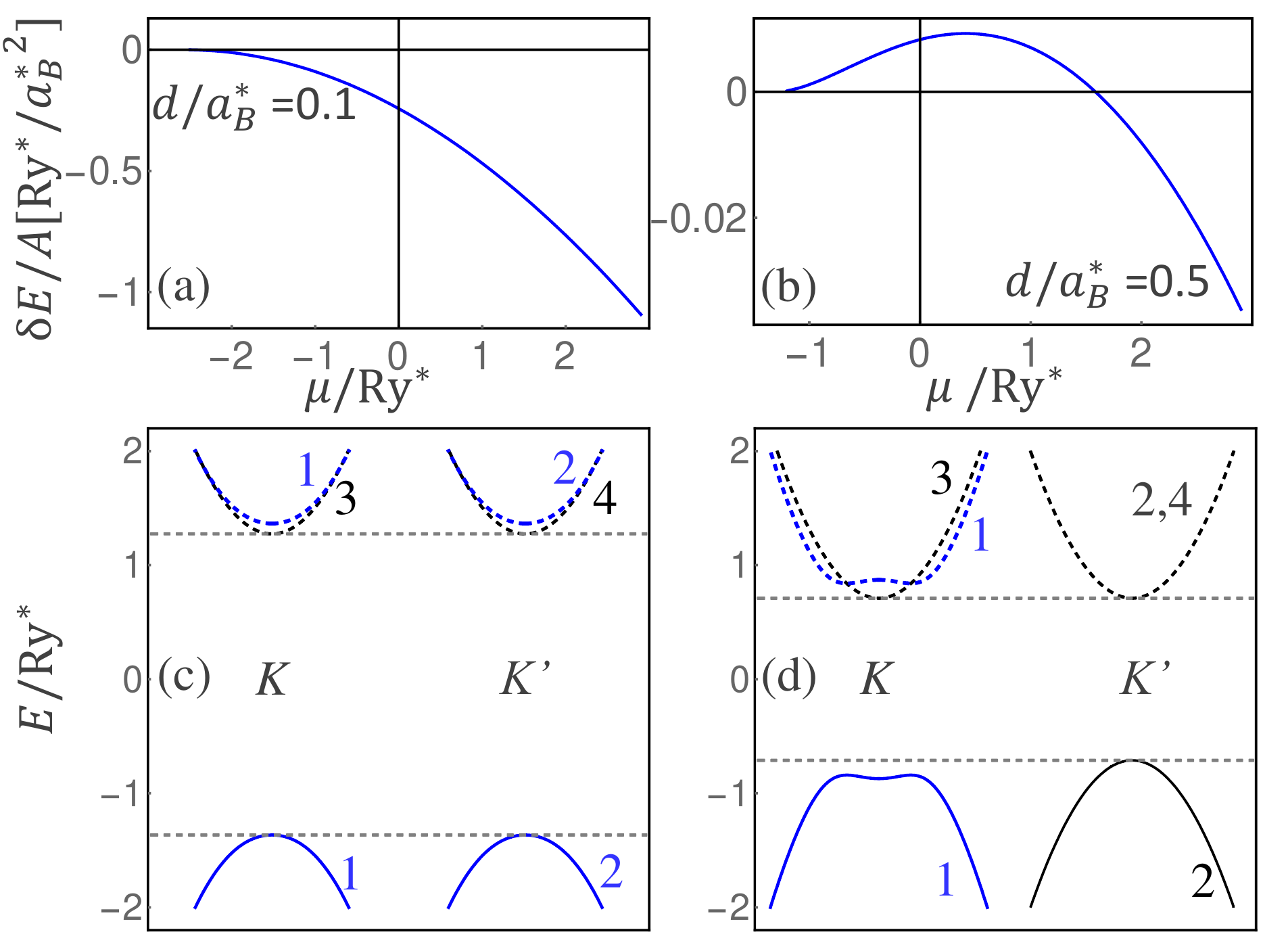}
	\caption{(Color online)
(a)-(b) Energy difference ($\delta E \equiv E_{\text{II}}-E_{\text{I}}$) per area between phase-I and II states 
as a function $\mu$ at $d/a_B^*=0.1$ (a) and $d/a_B^*=0.5$ (b).
(c)-(d) Typical quasiparticle energy bands in phase-II (c), and in phase-I (d) with 
(blue) and without (black) interlayer coherence.
$(d/a_B^*,\mu/Ry^*)$ is (0.1,-2.5) in (c) and (0.5,-1.1) in (d), corresponding to the two 
crosses($\times$) in Fig.~\ref{Fig:PhaseDiagram}. These results were calculated 
with $m_e=m_h$. 
}
	\label{Fig:Energy}
\end{figure}

\begin{table}[t]
	\caption{Parameter values for different combinations of monolayer layer 2H-TMDs. $m_0$ is the electron bare mass. $\epsilon=5$ for hBN. Electrons reside in MoS$_2$, and holes in other TMDs.  The listed energy gaps $E_g$ apply in the 
	absence of a gate voltage.}
	\resizebox{\columnwidth}{!}{
		\begin{tabular}{ l | c | c | c |c | c}
			\hline
			& $m_e/m_0$[\onlinecite{mass2015}] & $m_h/m_0$[\onlinecite{mass2015}] & $a_B^*$(\AA) & $\text{Ry}^*$(meV) & $E_{g}$(eV)[\onlinecite{Cho2013}]\\
			\hline
			MoS$_2$/MoTe$_2$ & 0.47  & 0.62 & 9.89 & 145 & 1.1  \\
			\hline
			MoS$_2$/WSe$_2$ & 0.47 & 0.36 & 12.97 & 111 & 1.4  \\
			\hline
			MoS$_2$/WTe$_2$ & 0.47 & 0.32 & 13.89 & 104 & 0.8 \\
			\hline
		\end{tabular}
	}
	\label{Parameters}
\end{table}

The indirect exciton binding energy $E_b$ determines the value for $\mu$ at which 
excitons first appear.  When $\mu < -E_b$ no excitons are present.  In this state each layer is electrically neutral and 
there is no interlayer coherence.  Eq.~(\ref{SC}) has nontrivial 
($n, \Delta_{\vec{k}} \neq 0$) solutions only for $\mu >-E_b$.
We find two distinct types of BXC phase. In phase-I, only one type of exciton condenses ({\it e.g.} $\Delta_{\vec{k}}^{(11)} \neq 0$ 
and $\Delta_{\vec{k}}^{(22)}=0$).  In phase-II, excitons associated with both valence bands
condense and have equal population
({\it  e.g.} $\Delta_{\vec{k}}^{(11)}=\Delta_{\vec{k}}^{(22)} \neq 0$).
Both phases are allowed by Eq.~(\ref{SC}).
We obtain the phase diagram in Fig.~\ref{Fig:PhaseDiagram}
by comparing the total energy of phase-I and II
as a function of $(d, \mu)$.  Below a critical layer separation 
$d_c\approx 0.25 a_B^*$, phase-II always has a lower energy, as illustrated in Fig.~\ref{Fig:Energy}(a). 
Above $d_c$, a transition from phase-I to phase-II occurs as the chemical potential $\mu$ increases(Fig.~\ref{Fig:Energy}(b)). 
Typical quasiparticle energy bands in phase-II and I are depicted in Fig.~\ref{Fig:Energy}(c) and (d), 
and show that the system is an excitonic insulator with a charge gap.

In our mean-field theory, condensation of one type or the other always occurs at 
$T=0$ when excitons are present.  It is well known however that at high electron and 
hole densities a first-order Mott transition occurs \cite{Liu1998,DePalo2002,Nikolaev2008,Asano2014,Fogler2014}
from the gapped exciton condensate phase to an ungapped electron-hole plasma state.  The electron-hole plasma 
state is preferred energetically because it can achieve better correlations between like-charge particles, reducing the 
probability that they are close together, while maintaining good correlations between oppositely-charged particles. 
The density at which the Mott transition occurs is most reliably estimated via a non-perturbative approaches.\cite{DePalo2002} 
No estimate is currently available for the TMD case, for which the valley degeneracy and the small spin-splitting in the conduction 
band will tend to favor plasma states over exciton condensate states.  Based on existing estimates\cite{DePalo2002} we can conclude that the 
Mott transition density is below $ n a_B^{*2} \sim 0.3$ as $d /a_B^* \to 0$ and below 
$ n a_B^{*2} \sim 0.05$ for $d/a_B^* \sim 1$.
Corrections to mean-field theory which go in the direction of favoring plasma states
can be partially captured by accounting for screening of the electron-hole interaction
which becomes stronger as exciton sizes increase and excitons correspondingly become
more polarizable.  The results reported here are intended to be reliable only in the low 
exciton density limit.   

\section{Interacting Boson Model}
\label{BoseM}
To understand the phase diagram more deeply, 
we employ a boson Hamiltonian designed to describe weakly-interacting excitons in the low density limit.  
Our strategy to obtain the boson Hamiltonian is to construct a Lagrangian based on a al wavefunction which
parametrizes a family of states with electron-hole coherence. The Berry phase part of the Lagrangian has the same form as that in the field-theory functional integral representation of a standard interacting boson model.\cite{Negele1988} 
Appealing to this property, we promote variational parameters in the wavefunction to bosonic operators. 
The details of the derivation are presented in Appendices~\ref{sec.BModel} and \ref{App2}.

The boson Hamiltonian is:
\begin{equation}
\begin{aligned}
&H_B=\sum(\frac{\hbar^2Q^2}{2M}-E_b-\mu)B_{(vc)\vec{Q}}^\dagger B_{(vc)\vec{Q}}\\
&+\frac{1}{2A}\sum' \big\{g_H(\vec{Q}_{14})B_{(vc)\vec{Q}_1}^\dagger B_{(v' c')\vec{Q}_2}^\dagger B_{(v' c')\vec{Q}_3}B_{(vc)\vec{Q}_4}\\
&+ g_X(\vec{Q}_{13},\vec{Q}_{14})B_{(vc)\vec{Q}_1}^\dagger B_{(v' c')\vec{Q}_2}^\dagger B_{(v' c)\vec{Q}_3}B_{(v c')\vec{Q}_4}
\big\},
\end{aligned}
\label{HBoson0}
\end{equation} 
where $B_{(vc)\vec{Q}}$ is a bosonic operator for an exciton
with a hole in valence band $v$, an electron in conduction band $c$,
and total momentum $\vec{Q}$. $\vec{Q}_{ab}$ is the momentum transfer $\vec{Q}_{a}-\vec{Q}_{b}$.  
For the TMD system, there are 8 possibilities for the composite index $(vc)$.
The quadratic term in Eq.~(\ref{HBoson0}) accounts for exciton kinetic energy ($M=m_e+m_h$) and chemical 
potential. The quartic terms describe exction-exciton interactions. 
The prime on the quartic term summation enforces momentum conservation $\vec{Q}_1+\vec{Q}_2=\vec{Q}_3+\vec{Q}_4$.

\begin{figure}[t]
	\vspace{+0.0cm}
	\includegraphics[width=1.0\columnwidth]{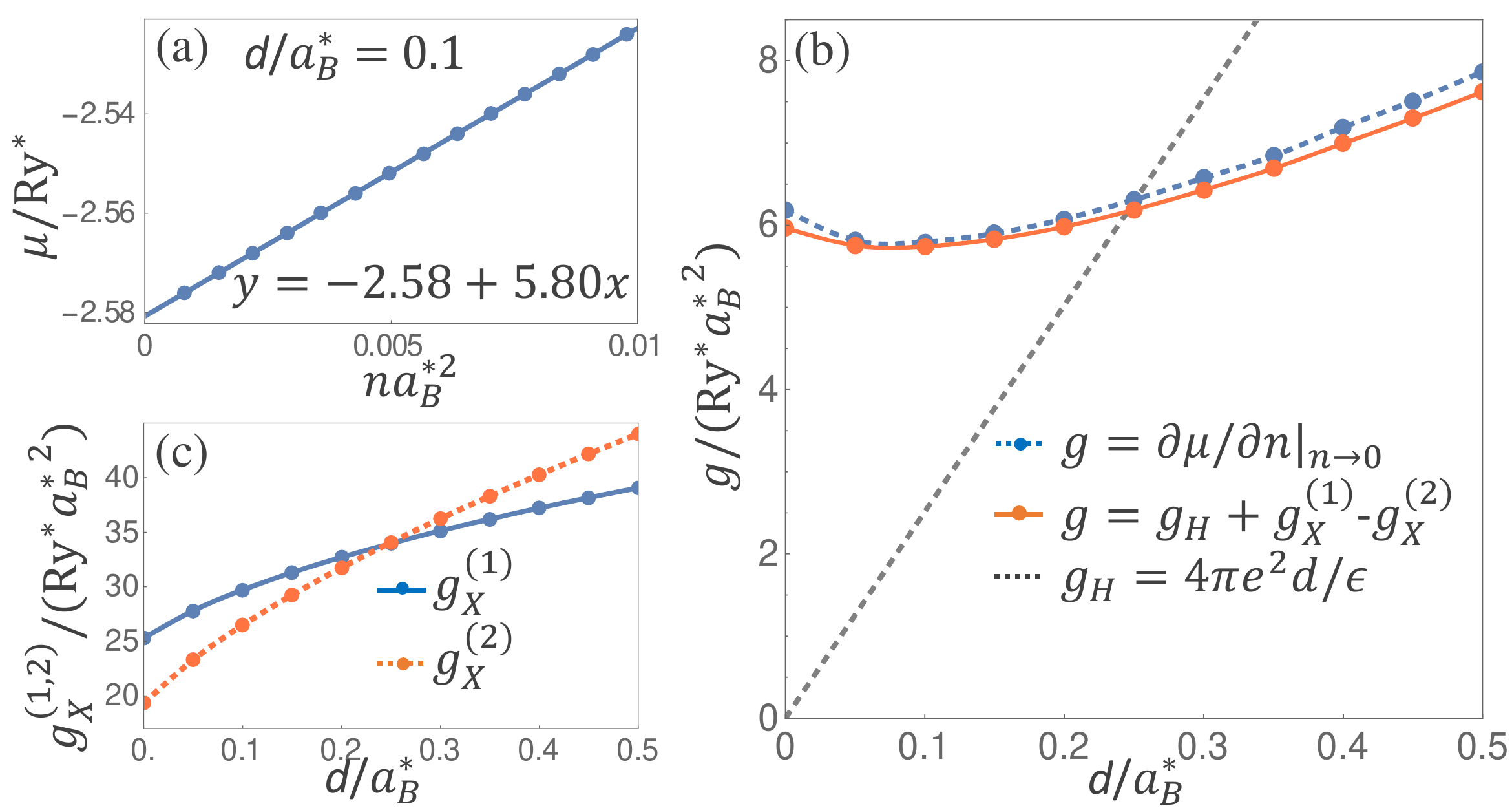}
	\caption{(Color online)(a)Chemical potential $\mu$ as a function of density $n$ obtained from Hartree-Fock calculations 
	in which a single exciton flavor is condensed. The line is a linear fit of the numerical data to Eq.~(\ref{mu_n}). 
	(b)Exciton-exciton interaction strength extracted from the self-consistent Hartree-Fock equation solutions (blue dashed line), and 
	exciton-exciton interaction strength calculated from the interacting boson model (red solid line). 
	The black dashed line is the Hartree contribution $g_H$ to the total interaction strength.  
		(c)The blue and red lines separate the interlayer ($g_X^{(1)}$) and intralayer ($g_X^{(2)}$) exchange contributions
		to the total exciton-exciton interaction strength.}
	\label{Fig:Interactionvsd}
\end{figure}

The two types of exciton interaction arise from the fermionic Hartree
and exchange interactions respectively. 
In the exchange interaction, two excitons swap constituent electrons or holes.
Analytic expressions for the coupling strength $g_H(\vec{q})$ and $g_X(\vec{q}~', \vec{q})$ are given in Appendix~\ref{sec.BModel}.

We focus here on their zero-momentum limits $g_H=g_H(0)$ and $g_X=g_X(0, 0)$, which are more easily 
interpreted and capture much of the exciton-exciton interaction physics. 
For the case in which the exciton condensate is populated by a single flavor we find that for low exciton densities 
\begin{equation}
\mu=-E_b+gn.
\label{mu_n}
\end{equation}
where $g=g_H+g_X$ is the total exciton-exciton interaction, as expected from the mean-field theory for weakly interacting bosons.
This behavior is illustrated in Fig.~\ref{Fig:Interactionvsd}(a).  We have verified that 
the interaction parameter obtained by examining the dependence of $\mu$ on $n$ in the fermion 
mean-field theory agrees with the analytic expression in App.~\ref{sec.BModel}, as illustrated in 
Fig.~\ref{Fig:Interactionvsd}(b) which plots $g$ as a function of layer separation $d$. 
We find that $g_H=4\pi e^2 d /\epsilon$ and that $g_X=g_X^{(1)}-g_X^{(2)}$, where
$g_X^{(1)}$ and $g_X^{(2)}$ are both positive and originate from inter and intra layer fermionic exchange interactions 
respectively. 
The binding energy of isolated excitons is due microscopically to attractive inter layer exchange interactions.  When excitons overlap and interact with each other, coherence between layers is 
reduced weakening inter layer exchange, but strengthening intra layer exchange.  This explains the signs
of the two contributions to $g_X$.  
The overall sign of $g_X$ is positive at $d=0$ because the loss of interlayer exchange energy
when excitons overlap is greater than the gain in intralayer exchange energy.  In Fig.~\ref{Fig:Interactionvsd}(c),
we show that $g_X$ becomes negative beyond a critical layer separation $d_c \sim 0.25a_B^*$.
It turns out that although both $g_X^{(1)}$ and $g_X^{(2)}$ increase with layer separation $d$,
the rate of increase of $g_X^{(1)}$ is smaller than for $g_X^{(2)}$. 
The difference in behavior can be traced to the exponential decrease in the momentum 
space inter-layer Coulomb interaction with layer separation $d$ as shown in Eq. ~(\ref{gX1}) and ~(\ref{gX2}). 

To find the ground state in the realistic multi-flavor case, we assume that all excitons condense
into $\vec{Q}=0$ states and introduce the following matrix:
\begin{equation}
\mathcal{F}=\frac{1}{\sqrt{A}}\begin{pmatrix}
\langle B_{(11)} \rangle & \langle B_{(12)} \rangle & \langle B_{(13)} \rangle & \langle B_{(14)} \rangle \\
\langle B_{(21)} \rangle & \langle B_{(22)} \rangle & \langle B_{(23)} \rangle & \langle B_{(24)} \rangle 
\end{pmatrix}.
\end{equation}
Neglecting the small spin-orbit splitting of conduction band states, 
the total energy per area can be written in a compact form,
\begin{equation}
\label{eq:totalenergy}
\frac{\langle H_B \rangle}{A}=-(E_b+\mu)\text{Tr}\mathcal{T}
+\frac{g_H}{2}(\text{Tr}\mathcal{T})^2
+\frac{g_X}{2}\text{Tr}\mathcal{T}^2,
\end{equation}
where $\mathcal{T}=\mathcal{F}^\dagger\mathcal{F}$.
In Eq.~(\ref{eq:totalenergy}) $\text{Tr}\mathcal{T}$ is the total density of excitons,summed over all flavors,and $\text{Tr}\mathcal{T}^2 - (\text{Tr}\mathcal{T})^2$ measures the 
flavor polarization of the exciton condensate.
This energy functional is invariant under the following transformation:
\begin{equation}
\mathcal{F}\mapsto e^{i\theta}\mathcal{U}_2\mathcal{F}\mathcal{U}_4^\dagger.
\end{equation}
Here $e^{i\theta}$ captures the U(1) symmetry which originates from separate charge conservation in the individual layers.
$\mathcal{U}_2$ and $\mathcal{U}_4$ are respectively $2\times 2$ and $4\times 4$ special unitary matrices, 
which capture the SU(2) symmetry of the valence bands and the SU(4) symmetry present in the conduction bands
when their spin-splitting is neglected.  The overall symmetry group 
of the system is U(1)$\times$SU(2)$\times$SU(4).
When the conduction band spin-orbit splitting is included, the higher energy conduction band states in each valley are not occupied and the symmetry group 
is reduced to U(1)$\times$SU(2)$\times$SU(2), corresponding to 
separate charge conservation and rotations in both conduction and valence band 
valley spaces.
 
$\mathcal{F}$ acquires a nonzero value in the ground state only if $\mu>-E_b$, . 
By minimizing the energy functional, we verify that the sign of $g_X$ 
determines the position of a phase boundary between two different classes of exciton condensate which 
we refer to as phase-I and II.  When $g_X<0$, phase-I is energetically favorable and a representative
realization of the ground state is,
\begin{equation}
\mathcal{F}_\text{I}=\sqrt{n_\text{I}}\begin{pmatrix}
1 & 0  & 0 & 0 \\
0 & 0  & 0 & 0 
\end{pmatrix},
\label{FI}
\end{equation}
where $n_\text{I}=(\mu+E_b)/(g_H+g_X)$ is the exciton density. $\mathcal{F}_\text{I}$ is invariant under the transformation:
\begin{equation}
\begin{pmatrix}
e^{i\phi} & 0 \\
0 & e^{-i\phi}  
\end{pmatrix}
\mathcal{F}_\text{I}\begin{pmatrix}
e^{-i\phi} & 0 \\
0 & \mathcal{V}_3^\dagger 
\end{pmatrix} =\mathcal{F}_\text{I},
\end{equation}
where $\mathcal{V}_3$ is a $3\times 3$ unitary matrix. 
Therefore, phase-I spontaneously breaks the U(1)$\times$SU(2)$\times$SU(4) symmetry down to U(1)$\times$U(3) symmetry.

\begin{figure}[t]
	\includegraphics[width=1\columnwidth]{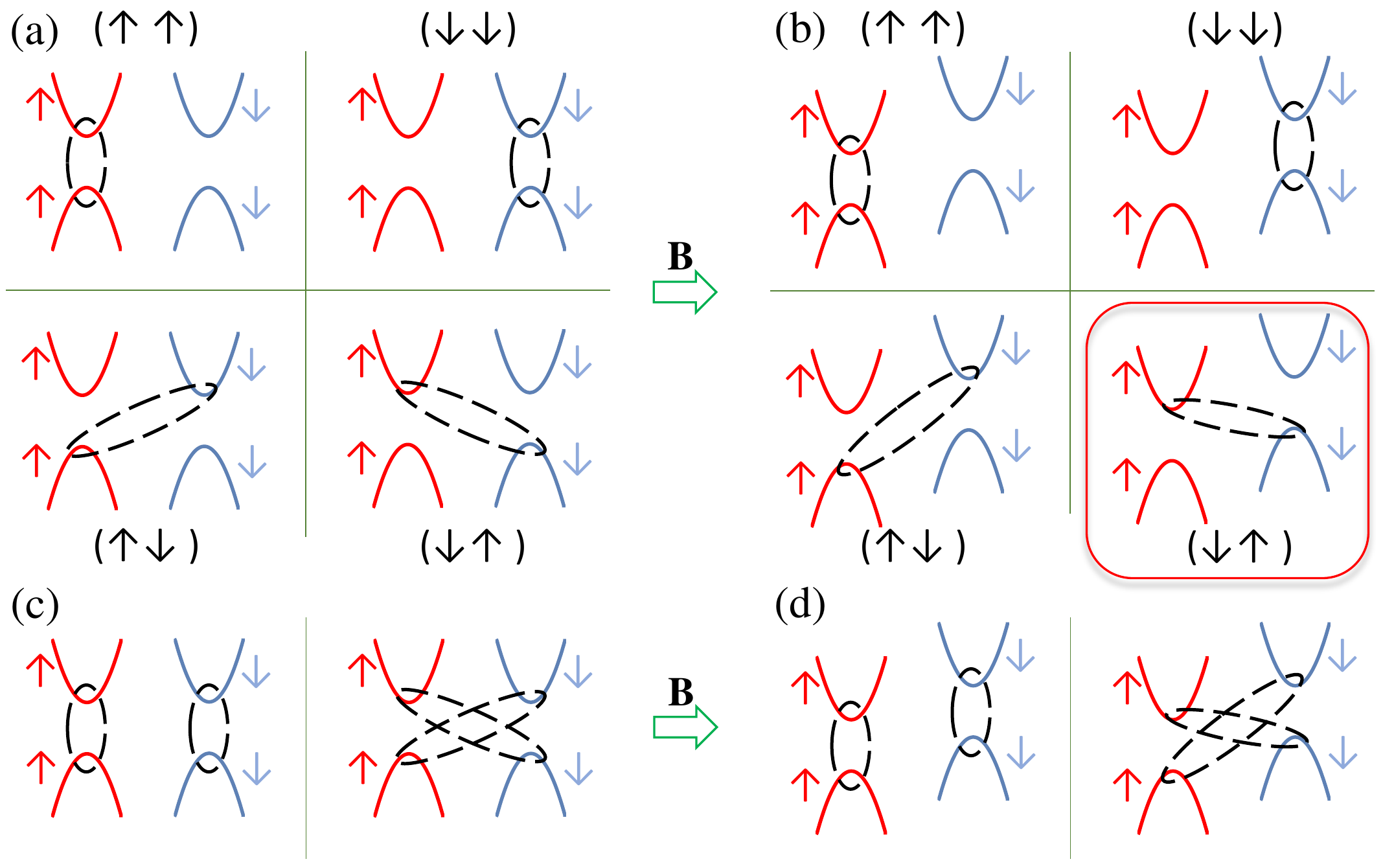}
	\caption{(Color online) (a) Four representative degenerate ground states in phase-I. For illustration purpose, only two conduction bands are shown. (The other two are slightly higher in energy because of spin-orbit splitting and do not participate in the ground state manifold.)
	The dashed black oval highlights the two bands with spontaneous phase coherence in phase-I. Charge is transferred from the valence band partner to the conduction band partner. In the upper panels $(\uparrow \uparrow)$ and $(\downarrow \downarrow)$, coherence is established between like spins and the ground state is not spin-polarized.  In the lower panels $(\uparrow \downarrow)$ and $(\downarrow \uparrow)$, charge is transferred between opposite spins and the ground state is spin-polarized, with a finite spin-polarization that is proportional to the transfered charge density.  Because spin and valley are locked by spin-orbit coupling, spin-polarized states are stabilized by an infinitesimal Zeeman field.	
	(b) In the presence of an infinitesimal Zeeman field that favors spin up, the spin-polarized state $(\downarrow \uparrow)$ is selected as the unique ground state in phase-I. (c) Two representative degenerate ground states in phase-II. Both states have zero spin-polarization, and remain degenerate in an infinitesimal Zeeman field as shown in (d).}
	\label{Fig:FM}
\end{figure}

When $g_X>0$ phase-II is realized.  Energy minimization shows that a 
representative realization of the ground state in phase-II is,
\begin{equation}
\mathcal{F}_\text{II}=\sqrt{n_\text{II}/2}\begin{pmatrix}
1 & 0  & 0 & 0 \\
0 & 1  & 0 & 0
\label{FII} 
\end{pmatrix},
\end{equation}
where $n_\text{II}=(\mu+E_b)/(g_H+g_X/2)$ is the total exciton density in phase-II. $\mathcal{F}_\text{II}$ is invariant under the transformation:
\begin{equation}
\mathcal{U}_2\mathcal{F}_\text{II}\begin{pmatrix}
\mathcal{U}_2^\dagger & 0 \\
0 & \mathcal{V}_2^\dagger 
\end{pmatrix} =\mathcal{F}_\text{II},
\end{equation}
where $\mathcal{V}_2 $ is a $2\times 2$ unitary matrix. 
Phase-II spontaneously breaks the U(1)$\times$SU(2)$\times$SU(4) symmetry down to SU(2)$\times$U(2) symmetry.
A similar analysis can be applied to identify the symmetry breaking pattern 
when the spin splitting of the conduction bands is considered.
In phase-I, an application of an infinitesimal external Zeeman field lifts both 
conduction and valence band valley degeneracies, and 
selects a unique condensate ground state with a finite 
spin-polarization, as illustrated in Fig.~\ref{Fig:FM}. 
Phase-I therefore satisfies the definition of a ferromagnet, defined as a system with a finite spin-polarization
in an infinitesimal Zeeman field, but has a distinct set of broken symmetries compared to the usual spin rotational symmetry breaking.

Based on this mean-field calculation, we conclude that although the system has
8 types of excitons in total, only one or two flavors condense in the ground state. 
The number of condensed flavors is in general limited by the number of distinct valence or 
conduction bands, which ever is smaller in number.  
Although the boson model correctly captures the phase transition position as a function of $d$, it is 
important to emphasize that it is valid only in the low exciton density limit.  
For this reason, it fails to accurately predict the $\mu$ dependence of the 
phase boundary. 
In addition it fails to capture the tendency 
toward weaker electron-hole pairing at high exciton densities, which
eventually leads to an electron-hole quantum liquid state with 
no interlayer coherence.

The relationship between the exciton density $n$, the exciton chemical potential $\mu$ and the 
coupling strength $g$ makes it possible to extract the value of $g$ from capacitance measurement. 
The differential capacitance per area for the heterostructure is:
\begin{equation}
\mathcal{C}=e^2\frac{\partial n}{\partial \mu}.
\end{equation}
The Hartree coupling strength $g_H$ can be identified as the inverse of the geometric capacitance:
\begin{equation}
\mathcal{C}_\text{geo}=e^2/g_H=\epsilon/(4\pi d).
\end{equation}
Therefore, capacitance measurement provides a simple way to determine the value of $g_X$ in the low-exciton density limit:
\begin{equation}
e^2(\mathcal{C}^{-1}-\mathcal{C}_\text{geo}^{-1})=\begin{cases} 
      g_X<0, & \text{phase-I} \\
      \frac{1}{2}g_X>0, & \text{phase-II}
   \end{cases}.
\end{equation}
The sign of $g_X$ helps to distinguish phase-I and II.

\section{Fluctuations and Stability}
\label{Fluc}
The bilayer exciton condensate is a state with spontaneously broken continuous 
symmetries, and therefore hosts low-energy collective fluctuations.
Theoretical studies of fluctuation properties are of interest in part because 
they can reveal mean-field state\cite{Fedorov2014} instabilities.
The collective modes can be studied using the interacting boson model, which is described in detail in Appendix~\ref{CModeBoson}.
The interacting boson model admits analytic solutions for collective modes associated with exciton density, phase and flavor fluctuations in both phase-I and II. However, it is valid only in the low exciton density limit. 
Here, we study another approach that can be applied to any exciton density.
This approach is based on 
the following variational wave function which captures exciton density and phase fluctuations 
in phase-I:
\begin{equation} 
	|\Phi\rangle = \prod_{\vec{k}}\Big[\mathcal{Z}_{\vec{k}}+\sum_{\vec{Q}}z_{\vec{k}}(\vec{Q})\gamma_{(\vec{k}+\vec{Q}),1}^{\dagger}\gamma_{\vec{k},0}\Big]|XC\rangle.
\end{equation} 
where $\Ket{XC}$ is the phase-I ground state.
$\gamma_{\vec{k},0}^{\dagger}$ and $\gamma_{\vec{k},1}^{\dagger}$ are respectively quasiparticle creation operators for occupied and empty
quasiparticle states in $\Ket{XC}$ associated with the ground state condensate and are defined as follows: 
\begin{equation}
\begin{aligned}
\gamma_{\vec{k},0}^{\dagger}=&u_{\vec{k}}a_{c\vec{k}}^{\dagger}+v_{\vec{k}}a_{v\vec{k}}^{\dagger},\\
\gamma_{\vec{k},1}^{\dagger}=&v_{\vec{k}}a_{c\vec{k}}^{\dagger}-u_{\vec{k}}a_{v\vec{k}}^{\dagger},
\end{aligned}
\end{equation}
where $u_{\vec{k}}$ and $v_{\vec{k}}$ are parameters determined by self-consistent Hatree-Fock equations~(\ref{SC}),
\begin{equation}
u_{\vec{k}}=\sqrt{\frac{1}{2}(1-\xi_{\vec{k}}/E_{\vec{k}})},\quad
v_{\vec{k}}=\sqrt{\frac{1}{2}(1+\xi_{\vec{k}}/E_{\vec{k}})}. 
\label{uv}
\end{equation} 
$\mathcal{Z}_{\vec{k}}$ is a normalization factor,
\begin{equation}
\mathcal{Z}_{\vec{k}}=\sqrt{1-\sum_{\vec{Q}}|z_{\vec{k}}(\vec{Q})|^2},
\end{equation} and $z_{\vec{k}}(\vec{Q})$ are complex parameters.  

\begin{figure}[t]
	\includegraphics[width=1.0\columnwidth]{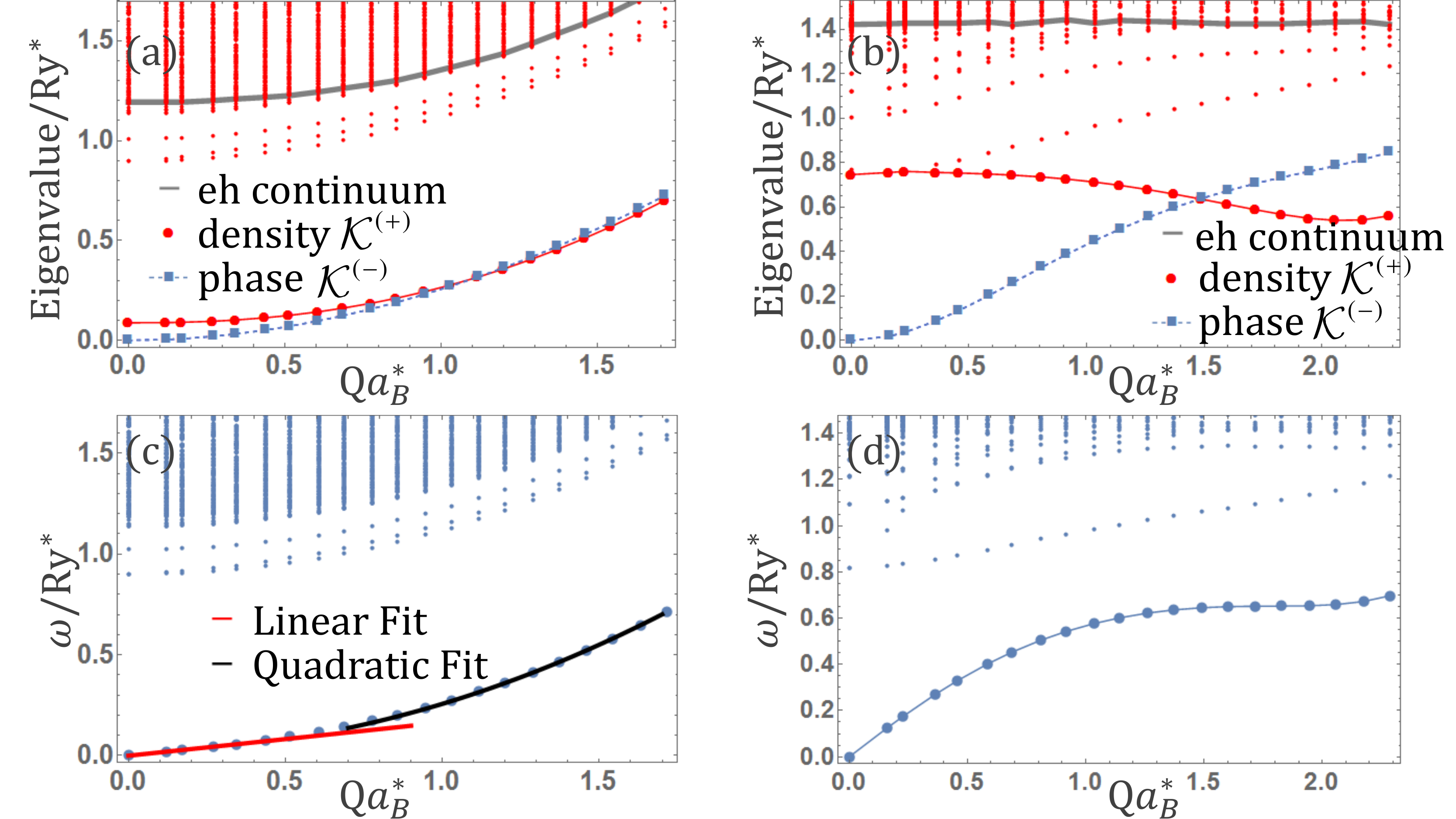}
	\caption{
	(Color online)(a)-(b) Eigenvalues of $\mathcal{K}^{(\pm)}$ for $d/a_B^*=0.5$ as a function of momentum $\vec{Q}$ at
	 $na_B^{*2}=0.008$ (a) and $na_B^{*2}=0.08$ (b). Red solid and blue dashed lines connect the lowest eigenvalues of 
	the density fluctuation matrix $\mathcal{K}^{(+)}$ and the phase fluctuation matrix $\mathcal{K}^{(-)}$ respectively. 
	Small red points label higher eigenvalues of $\mathcal{K}^{(+)}$.  The gray lines mark the lower edge of 
	the quasiparticle electron-hole continua [$\text{Min}(E_{\vec{k}}+E_{\vec{k}+\vec{Q}})$]. 
	(c)-(d) Collective mode spectra at $na_B^{*2}=0.008$ (c) and $na_B^{*2}=0.08$ (d) for $d/a_B^*=0.5$. 
	Larger points are used for the lowest energy excitations.  
	In (c), the red and black lines are respectively a linear fit at small $Q$ and a quadratic fit at large $Q$ to the 
	collective mode energy. For the results presented here, we assumed $m_e=m_h$.}
	\label{Fig:Stability and Fluctuation}
\end{figure}

To study fluctuation dynamics, we construct the Lagrangian:
\begin{equation}
\mathcal{L}=\langle \Phi | i \partial_t -H|\Phi \rangle \approx \mathcal{B}-\delta E^{(2)},
\label{Lfluc}
\end{equation}
where $\mathcal{B}$ is the harmonic Berry phase, and $\delta E^{(2)}$ is the harmonic energy variation \cite{Giuliani2005}:
	\begin{equation}
	\begin{aligned}
	&\delta E^{(2)}=\sum_{\vec{Q},\vec{k},\vec{p}}\{\mathcal{E}_{\vec{k},\vec{p}}(\vec{Q})z_{\vec{k}}^*(\vec{Q})z_{\vec{p}}(\vec{Q})\\
	&+\frac{1}{2}\Gamma_{\vec{k},\vec{p}}(\vec{Q})
	[z_{\vec{k}}(\vec{Q})z_{\vec{p}}(-\vec{Q})+z^*_{\vec{k}}(\vec{Q})z^*_{\vec{p}}(-\vec{Q})]\}.
	\end{aligned} 
		\label{EF}
	\end{equation}
Explicit forms for the matrices $\mathcal{E}$ and $\Gamma$ are given in App.~\ref{App3}.
To decouple $\pm\vec{Q}$ contributions in Eq.~(\ref{EF}), we perform a change of variables,
defining 
\begin{align}
z_{\vec{k}}(\vec{Q})&=x_{\vec{k}}(\vec{Q})+\mathrm{i}y_{\vec{k}}(\vec{Q})\\ 
z^*_{-\vec{k}}(-\vec{Q})&=x_{\vec{k}}(\vec{Q})-\mathrm{i}y_{\vec{k}}(\vec{Q}).
\end{align} 
Note that $x_{\vec{k}}(\vec{Q})$ and $y_{\vec{k}}(\vec{Q})$ are complex numbers, and that there is a redundancy,
\begin{equation}
\begin{aligned}
x_{-\vec{k}}(-\vec{Q})&=x^*_{\vec{k}}(\vec{Q}),\\
y_{-\vec{k}}(-\vec{Q})&=y^*_{\vec{k}}(\vec{Q}).
\end{aligned}
\label{redundancy}
\end{equation}	 
In terms of the $x$ and $y$ fields, the Berry phase $\mathcal{B}$ and energy variation $\delta E^{(2)}$ are,
\begin{equation}
\begin{aligned}
&\mathcal{B}=\frac{1}{2}\sum_{\vec{Q},\vec{k}}\Big(
y^*\partial_t x+y\partial_t x^*
-x^*\partial_t y-x \partial_t y^*
 \Big)_{\vec{k}}(\vec{Q}),\\
&\delta E^{(2)}=\sum_{\vec{Q}}\sum_{\vec{k},\vec{p}}\Big( x^*_{\vec{k}} \mathcal{K}^{(+)}_{\vec{k},\vec{p}} x_{\vec{p}}
+ y^*_{\vec{k}} \mathcal{K}^{(-)}_{\vec{k},\vec{p}} y_{\vec{p}}\Big)(\vec{Q}).
\end{aligned}
\end{equation}
The kernel matrices $\mathcal{K}^{(\pm)}_{\vec{k},\vec{p}}(\vec{Q})=\big(\mathcal{E}_{\vec{k},\vec{p}}\pm \Gamma_{\vec{k},-\vec{p}}\big)(\vec{Q})$ are real and symmetric. The $x$ and $y$ fields in $\delta E^{(2)}$ 
can be identified with exciton density and phase respectively, and the Berry phase contribution to
the action captures the conjugate relationship between these fluctuation variables.

Stability of the mean-field ground states against small fluctuations requires 
that the matrices $\mathcal{K}^{(\pm)}$ are nonnegative.
We have verified that this condition is satisfied out to large $d$ 
by explicit numerical calculations like those summarized in Fig.~\ref{Fig:Stability and Fluctuation}(a) and (b).
At $\vec{Q}=0$, the matrix $\mathcal{K}^{(-)}$ always has a zero-energy eigenvalue since,
\begin{equation}
\sum_{\vec{p}}\mathcal{K}^{(-)}_{\vec{k},\vec{p}}(0)\frac{\Delta_{\vec{p}}}{E_{\vec{p}}}=0,
\end{equation}
which follows from the fact that ground state energy is independent of global interlayer phase.  

For low exciton density $n$ (Fig.~\ref{Fig:Stability and Fluctuation}(a)), 
the lowest eigenvalues of $\mathcal{K}^{(+)}$ and $\mathcal{K}^{(-)}$ have similar behavior and are 
separated from the continuum. This is expected since $\mathcal{K}^{(+)}$ and 
$\mathcal{K}^{(-)}$ are identical 
in the limit  $n \rightarrow 0$. Fig.~\ref{Fig:Stability and Fluctuation}(b) demonstrates that the 
lowest eigenvalues of $\mathcal{K}^{(+)}$ are close to the particle-hole continuum when the exciton density becomes large;
the interacting boson model discussed above fails qualitatively in this limit.

The Euler-Lagrange equation for the Lagrangian in Eq.~(\ref{Lfluc}) gives rise to the equation of motion,
\begin{equation}
\begin{aligned}
\partial_t x_{\vec{k}}(\vec{Q})=-\Big(\mathcal{K}^{(-)}_{\vec{k},\vec{p}} y_{\vec{p}}\Big)(\vec{Q}),\\
\partial_t y_{\vec{k}}(\vec{Q})=+\Big(\mathcal{K}^{(+)}_{\vec{k},\vec{p}} x_{\vec{p}}\Big)(\vec{Q}).
\end{aligned}
\end{equation}
which leads to
\begin{equation}
\partial_t^2 y_{\vec{k}}(\vec{Q})=
-\Big[\big(\mathcal{K}^{(+)}\mathcal{K}^{(-)}\big)_{\vec{k},\vec{p}} y_{\vec{p}}\Big](\vec{Q}).
\end{equation}
It follows that the energy of the collective mode is given by the square root of the lowest eigenvalues of the matrix product $\mathcal{K}^{(+)}\mathcal{K}^{(-)}$, which is plotted in Fig.~\ref{Fig:Stability and Fluctuation}(c) and (d).
The lowest energy collective mode is the gapless Goldstone mode of the exciton condensate. 
For low exciton density $n$ (Fig.~\ref{Fig:Stability and Fluctuation}(c)), the Goldstone mode has linear dispersion at small 
$\vec{Q}$, becoming quadratic at large $\vec{Q}$. 
This agrees with the Goldstone mode behavior predicted by the weakly interacting boson model(Eq.~(\ref{HBoson0})). 
For large $n$ (Fig.~\ref{Fig:Stability and Fluctuation}(d)), the Goldstone mode deviates from 
quadratic behavior at large $\vec{Q}$.  The failure of the weakly interacting boson model in the high density limit 
originates from the internal structure of the excitons.  When the typical distance between excitons is
comparable to exciton size, excitations must 
be described in terms of the underlying conduction and valence 
band fermion states.\cite{Keldysh1965,Comte1982,Zhu1995}

\section{Summary and Discussion}
\label{summ}
By combining Hartree-Fock theory and an interacting boson model,
we have shown that spatially indirect exciton condensates in group-VI TMD bilayers have two distinct phases.
We have also studied the dynamics of exciton condensate density and phase fluctuations and 
calculated the associated collective mode spectra.

The topic of exciton condensation in semiconductors has a long history and 
our work is related to some earlier studies.  For example, 
Berman {\it et al.}\cite{Berman2012} studied exciton condensation in bilayers
formed from gapped graphene, although the possibility of two distinct condensate phases was not considered.
The phase transition between the two condensate phases as a function of layer separation was studied previously
\cite{Tejedor97,Laikhtman2001} for the case of quantum well bilayer excitons, and further 
explored in a very recent publication.\cite{Dubin2015}
The TMD layers considered in this paper are distinguished from semiconductor quantum well
systems by exciton binding energies that are an order of magnitude larger, and by
spin-valley coupling which leads to two-fold degenerate valence bands and 
approximately four-fold degenerate conduction bands.  
Compared to Refs.\onlinecite{Laikhtman2001} and \onlinecite{Dubin2015}, we used a completely different approach
to derive an interacting boson model. Our approach is physically transparent, and is based 
on a variatonal wavefunctions defined by parameters whose quantum fluctuations are 
characterized by using a Lagrangian formalism. 
The bosonic nature of excitons is automatically taken into account in the Lagrangian, and there is no need to calculate 
combinatorial factors arising from the indistinguishability of bosonic particles. 
Our approach provides a simple yet systematic way to model the exction-exciton interaction. 
We have also discussed a fermionic Hartree-Fock approach from which the exciton-exciton interaction strengths can be extracted
with similar results, and proposed that the interaction strengths can be experimentally determined by
performing capacitance measurement.

Because the hBN dielectric barrier in the systems of interest, must be thick enough to make interlayer tunneling weak,
Fig.~\ref{Fig:PhaseDiagram} implies that phase-I with a single condensate flavor is 
more likely to be realized in experiment than phase-II.  
Phase-I breaks the invariance of the system Hamiltonian under 
separate valley rotations in conduction and valence bands, and is ferromagnetic in the sense that infinitesimal Zeeman coupling leads to a spin-polarization that is 
proportional to the exciton density.

In spit of their large gaps, band edge states in TMDs have relatively large Berry curvatures\cite{Yao12}.
In monolayer TMDs momentum space Berry curvatures lead\cite{Wu2015} to unusual exictonic spectra
in which hydrogenic degeneracies are lifted.  Although band Berry curvatures should be less important in 
spatially indirect exciton systems because weaker binding implies that
the exciton states are formed within a smaller 
region of momentum space.  In terms of its influence on quasiparticle bands,
exciton condensation has the effect of preventing gaps between conduction and 
valence band states from closing.  Since the host semiconductor materials are 
topologically trivial, and since transitions between trivial and non-trivial states 
can occur continuously only when the quasiparticle charged excitation energy 
vanishes, we argue that
exciton condensation will not result in interaction-driven topologically nontrivial states in our system.

The critical temperature of spatially indirect exciton condensate is the 
Berezinskii-Kosterlitz-Thouless transition temperature,
given at low exciton densities by the weakly-interacting boson 
expression 
\begin{equation}
k_B T_{\text{BKT}}\approx 1.3 \frac{\hbar^2 n}{M}=2.6(na_B^{*2})\frac{m}{M}\text{Ry}^*, 
\end{equation}
where $M$ is the electron-hole pair total mass and $m$ is the reduced mass. 
In the low exciton density limit $k_B T_{\text{BKT}}$ scales linearly with 
exciton density.\cite{Bonitz2010,Fogler2014} 
For the MoS$_2$/hBN/MoTe$_2$ heterostructure and exciton densities $n$ in the  
$10^{12}$cm$^{-2}$ range, $na_B^{*2}\approx 0.01$ and $T_{\text{BKT}}$ is about 10K.  
The maximum possible transition temperature 
is closely related to the critical density at which the Mott transition to an 
electron-hole plasma occurs, and this increases as 
$d/a_{B}^*$ decreases.  Using the variational Monte Carlo estimate of 
DePalo {\it et al.}\cite{DePalo2002} the critical value of 
$n a_B^{*2} \sim 0.3$ as $d /a_B^* \to 0$.  From this we conclude 
that $k_B T_{\text{BKT}}$ cannot exceed around $300 K$.  
Adjustment of exciton density by external bias voltage can be employed to search for the highest transition 
temperature and to study the Mott transition to an ungapped electron-hole plasma that 
is expected at high exciton densities.  The most interesting regime 
is likely to be the case of very small layer separations of which current leakage 
driven through the tunnel barrier by an interlayer bias potential might be appreciable,
requiring the bilayer to be treated as a non-equilibrium system.     

The exciton condensate should be experimentally realizable in TMD bilayers 
provided that samples with sufficiently weak disorder can be achieved.
The photoluminescence line width $W$ of an individual monolayer 
TMD is a particularly useful characterization of 
sample quality for this purpose. 
$W$ is currently dominated\cite{Li2014} by the position-dependence of exciton 
energies. Therefore, the narrower the line width $W$, the weaker the disorder, and the better the sample quality.  
We expect bilayer exciton condensation to occur only in 
samples in which $W < k_B T_{\text{BKT}}$, since the excitons will otherwise simply localize near 
positions where they have minimum energy.  
Note that the inhomogeneous broadening $W$ of spatially indirect excitons will 
not be experimentally accessible since the corresponding transitions are optically inactive when the interlayer tunneling is negligible, 
but that it should be similar to the broadening of the readily measurable direct exciton energies.   
It should therefore be possible to judge on the basis of optical characterization 
when samples have achieved sufficient quality to study spatially indirect exciton condensate 
physics.  

\section{Acknowledgment}
This work was supported by the SRC and NIST under the Nanoelectronic Research Initiative (NRI)
and SWAN, and by the Welch Foundation under Grant No. F1473.

\appendix
\section{Interacting boson model for excitons in the low density limit}
\label{sec.BModel}

\begin{figure}[b]
	\includegraphics[width=1.0\columnwidth]{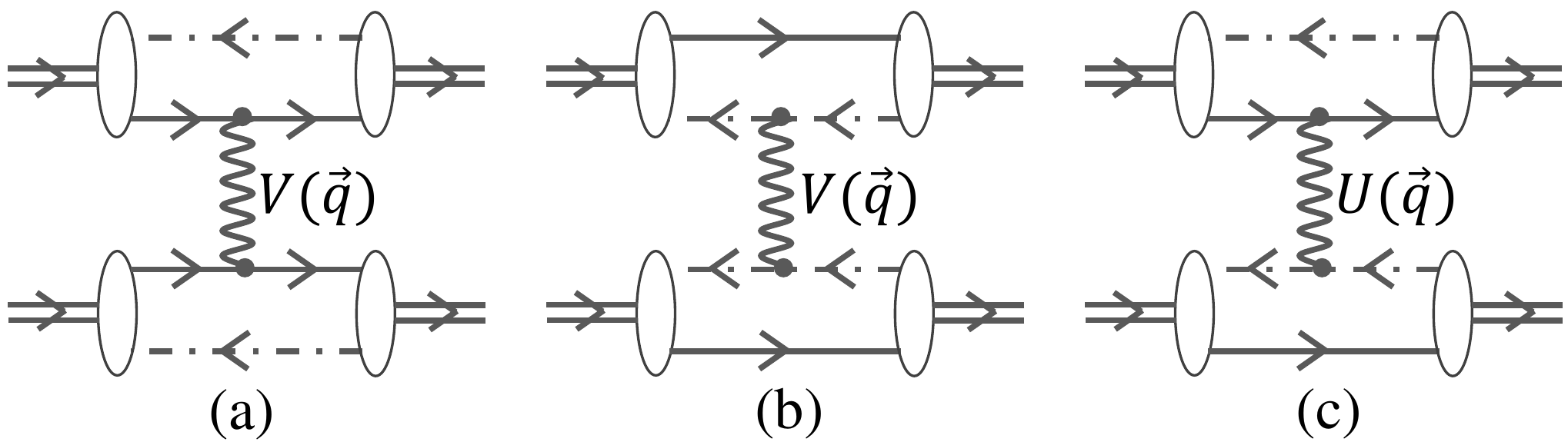}
	\caption{Feynman diagrams for the Hartree exciton-exciton interaction processes. A double line with arrow represents an 
	exciton state, and a single solid (dashed) line with arrow depicts an electron (hole) in the exciton. Wavy lines are interaction $V(\vec{q})$ or $U(\vec{q})$. (a) and (b) are the intralayer contributions, while (c) is the interlayer contribution. (a), (b) and (c) correspond to the three terms in $\mathcal{H}_H^{(4)}$ (Eq.~(\ref{En4})) and also the three terms in $g_H(\vec{q})$ (Eq.~(\ref{gH})).}
	\label{Fig:HI}
\end{figure}

\begin{figure}[b]
	\includegraphics[width=1.0\columnwidth]{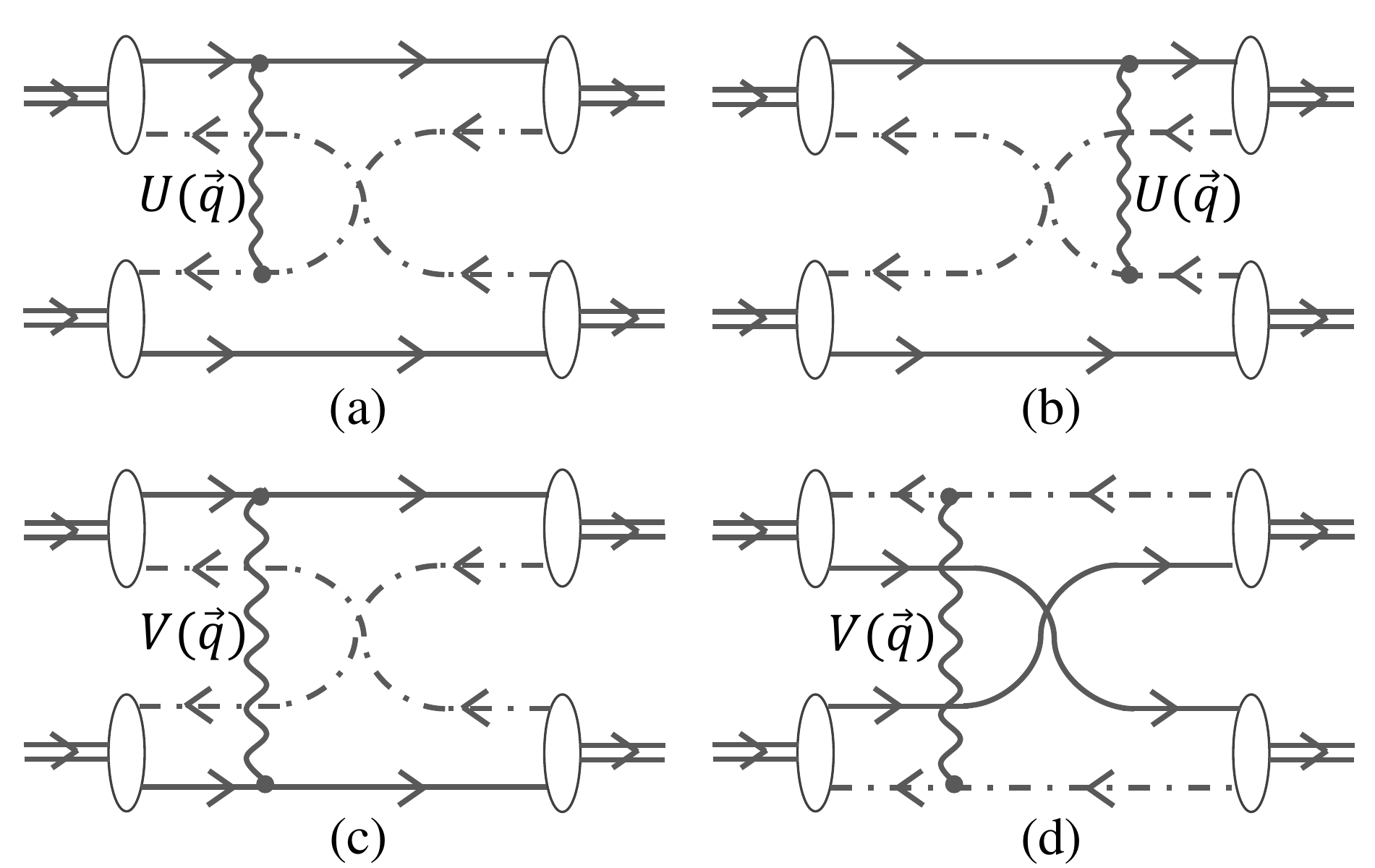}
	\caption{Feynman diagrams for the exchange exciton-exciton interaction processes. 
	The convention is the same as in Fig.~\ref{Fig:HI}.
		(a) and (b) are the interlayer contributions, which correspond to the first two terms in $\mathcal{H}_X^{(4)}$ (Eq.~(\ref{En4})) and  the two terms in  $g_X^{(1)}$ (Eq.~(\ref{gX1})). 
		(c) and (d) are the intralayer contributions, which correspond to the lase two terms in $\mathcal{H}_X^{(4)}$ (Eq.~(\ref{En4})) and  the two terms in  $g_X^{(2)}$ (Eq.~(\ref{gX2})). 
	}
	\label{Fig:XI}
\end{figure}

In the low density limit, excitons can be approximated as interacting bosons. 
We take a BCS like variational wave function to describe excitons, 
\begin{equation}
\begin{aligned}
|\Psi \rangle&=\frac{1}{\mathcal{N}}\exp(\Omega^\dagger)|\text{vac} \rangle,\\
\Omega^\dagger&=\sum_{V,C} \lambda_{VC} a_{C}^{\dagger} a_V,
\end{aligned}
\end{equation}
where $C$ and $V$ respectively denote a conduction and valence band state, and include internal indices such as spin and valley and also momentum label.
$|\text{vac}\rangle$ is the vacuum state defined by $a_V^\dagger|\text{vac}\rangle=a_C|\text{vac}\rangle=0$.
$\Omega^\dagger$ operator creates particle-hole excitations on top of the vacuum.
$\mathcal{N}$ is a normalization factor so that $\langle\Psi |\Psi \rangle=1$. 
$\lambda_{VC}$ is a set of complex variational parameters, which are small when the exciton density is low.

The density matrix with respect to $|\Psi \rangle$ is $\rho_{\alpha\beta}=\langle\Psi|a_\alpha^\dagger a_\beta|\Psi \rangle$,
where $\alpha$ and $\beta$ can be conduction or valence states.
We expand the density matrix to fourth order in $\lambda_{VC}$,
\begin{equation}
\begin{aligned}
\rho_{V V'}&\approx \delta_{V V'}+(-\lambda \lambda^\dagger+\lambda \lambda^\dagger \lambda \lambda^\dagger)_{V V'},\\
\rho_{C C'}&\approx(\lambda^\dagger \lambda-\lambda^\dagger \lambda \lambda^\dagger \lambda)_{C C'},\\
\rho_{V C}&\approx(\lambda-\lambda \lambda^\dagger \lambda)_{VC},
\end{aligned}
\end{equation}
where $\lambda$ is understood to be a matrix and $\lambda^\dagger$ is its Hermitian conjugate.

We introduce another matrix $\Lambda$ so that $\rho_{C C'}$ has a quadratic form without fourth-order correction,
\begin{equation}
\lambda=\Lambda+\frac{1}{2}\Lambda\Lambda^\dagger\Lambda.
\end{equation}
Expanding $\rho$ up to fourth order of $\Lambda$, we have that
\begin{equation}
\begin{aligned}
\rho_{V V'}&\approx\delta_{V V'}-(\Lambda \Lambda^\dagger)_{V V'},\\
\rho_{C C'}&\approx(\Lambda^\dagger \Lambda)_{C C'},\\
\rho_{V C}&\approx(\Lambda-\frac{1}{2}\Lambda \Lambda^\dagger \Lambda)_{VC}.
\end{aligned}
\end{equation}
The number of excitons is $\langle N_{ex} \rangle=\sum_{C}\rho_{CC}\approx\text{Tr}\Lambda^\dagger\Lambda$. 
Therefore, we verify that $\Lambda$ acts as the small parameter in the limit of low $\langle N_{ex} \rangle$.

An important property of the density matrix is that
\begin{equation}
\rho^2-\rho=O(\Lambda^5),
\label{r2r}
\end{equation}
which indicates that $|\Psi\rangle$ can be approximated as a Slater determinant up to fourth order in $\Lambda$.\cite{Giuliani2005}

$|\Psi \rangle$ parametrizes a family of states with electron-hole coherence, and also represents low-energy states in the low-exicton density limit. We choose to construct an effective interacting boson model using this variational wavefunction approach rather than a commonly used auxiliary field approach because of the necessity of consistently accounting for both exchange and Hartree mean-fields in spatially-indirect exciton systems. (See additional discussion below.)   
To study low-energy dynamics, we construct a Lagrangian based on $|\Psi \rangle$
\begin{equation}
\mathcal{L}=\langle \Psi| i \partial_t -H|\Psi \rangle=\mathcal{B}-\mathcal{H},
\label{Lag}
\end{equation}
and again expand everything to $\Lambda^4$.
This Lagrangian provides an effective field theory for excitons.  
The Berry phase has the following form,
\begin{equation}
\mathcal{B}=\langle \Psi| i \partial_t|\Psi \rangle \approx \frac{i}{2}\big(\text{Tr}[\Lambda^\dagger\partial_t\Lambda]
-\text{Tr}[(\partial_t\Lambda^\dagger)\Lambda]\big),
\label{BPhase}
\end{equation}
which does not have fourth order corrections.

To calculate the energy functional $\mathcal{H}$, we take advantage of the Slater determinant approximation\cite{Giuliani2005} to $|\Psi\rangle$ (Eq.~(\ref{r2r})) and obtain that 
\begin{equation}
\mathcal{H}=\langle \Psi| H |\Psi \rangle\approx \mathcal{H}^{(2)}+\mathcal{H}^{(4)}_H+\mathcal{H}^{(4)}_X,
\label{En}
\end{equation}
where $\mathcal{H}^{(2)}$ is quadratic in $\Lambda$, and $\mathcal{H}^{(4)}_{H,X}$ is quartic in $\Lambda$ with subscript $H$ and $X$ representing Hartree and exchange contributions. The explicit forms are below.
\begin{equation}
\begin{aligned}
\mathcal{H}^{(2)}=&\quad(\varepsilon_{C}-\varepsilon_{V})\Lambda^\dagger_{CV}\Lambda_{VC}\\
&-W_{V_1 C_1 C_2 V_2}\Lambda^\dagger_{C_1 V_2}\Lambda_{V_1 C_2},\\
\mathcal{H}^{(4)}_H=&\quad\frac{1}{2}W_{C_1 C_2 C_3 C_4}(\Lambda^\dagger\Lambda)_{C_1 C_4}(\Lambda^\dagger\Lambda)_{C_2 C_3}\\
&+\frac{1}{2}W_{V_1 V_2 V_3 V_4}(\Lambda\Lambda^\dagger)_{V_1 V_4}(\Lambda\Lambda^\dagger)_{V_2 V_3}\\
&-W_{V_1 C_1 C_2 V_2}(\Lambda\Lambda^\dagger)_{V_1 V_2}(\Lambda^\dagger\Lambda)_{C_1 C_2},\\
\mathcal{H}^{(4)}_X=&\quad\frac{1}{2}W_{V_1 C_1 C_2 V_2}\Lambda^\dagger_{C_1 V_2}(\Lambda\Lambda^\dagger\Lambda)_{V_1 C_2}\\
&+\frac{1}{2}W_{V_1 C_1 C_2 V_2}(\Lambda^\dagger\Lambda\Lambda^\dagger)_{C_1 V_2}\Lambda_{V_1 C_2}\\
&-\frac{1}{2}W_{C_1 C_2 C_3 C_4}(\Lambda^\dagger\Lambda)_{C_1 C_3}(\Lambda^\dagger\Lambda)_{C_2 C_4}\\
&-\frac{1}{2}W_{V_1 V_2 V_3 V_4}(\Lambda\Lambda^\dagger)_{V_1 V_3}(\Lambda\Lambda^\dagger)_{V_2 V_4}.
\end{aligned}
\label{En4}
\end{equation}
Here $\varepsilon_{C}$ and $\varepsilon_{V}$ are conduction and valence state energy including self-energy effects.
The interaction kernel $W$ has the form
\begin{equation}
\begin{aligned}
&W_{(n_1\vec{k}_1)(n_2\vec{k}_2)(n_3\vec{k}_3)(n_4\vec{k}_4)}\\
=&\frac{1}{A}\delta_{n_1 n_4}\delta_{n_2 n_3}
\delta_{\vec{k}_1+\vec{k}_2,\vec{k}_3+\vec{k}_4}W_{n_1 n_2}(\vec{k}_1-\vec{k}_4),
\end{aligned}
\end{equation}
where the momentum dependence is now explicit, and $n$ denotes internal indices.
$A$ is the area of the system. $W_{n_1 n_2}(\vec{q})$ is the intralayer interaction $V(\vec{q})$ if both $n_1$ and $n_2$ represent conduction or valence bands, and the interlayer interaction $U(q)$ otherwise.

We now write $\mathcal{H}^{(2)}$ in a more concrete form
\begin{equation}
\begin{aligned}
\mathcal{H}^{(2)}=&\Lambda_{(v,\vec{k})(c,\vec{k}+\vec{Q})}^*
\Big[\Big(\frac{\hbar^2(\vec{k}+\vec{Q})^2}{2m_e}+\frac{\hbar^2k^2}{2m_h}-\mu\Big)\delta_{\vec{k}\vec{k}'}\\
&-\frac{1}{A}U(\vec{k}-\vec{k}')\Big]\Lambda_{(v,\vec{k}')(c,\vec{k}'+\vec{Q})}.
\end{aligned}
\end{equation}
Here $v$ and $c$ denote different valence and conduction bands. We approximate $\varepsilon_{C}$ and $\varepsilon_{V}$ by parabolic bands, and assume different valence (conduction) bands have the same hole (electron) mass $m_h$ ($m_e$). In the case of TMDs, these are reasonable approximations, and $v$ and $c$ respectively take two and four different values. 

$\mathcal{H}^{(2)}$ can be reduced into a diagonal from by doing the following decomposition,
\begin{equation}
\Lambda_{(v,\vec{k})(c,\vec{k}+\vec{Q})}=\frac{1}{\sqrt{A}}f(\vec{k}+x_h \vec{Q}) B_{(vc)\vec{Q}}
\label{LB}
\end{equation}
where $B_{(vc)\vec{Q}}$ is a complex field that depends on momentum $\vec{Q}$ but not on $\vec{k}$.
$x_h=m_h/M$, where the total mass $M=m_e+m_h$. For notation convenience, we also introduce $x_e=m_e/M=1-x_h$.
$f(\vec{k})$ is the $1s$ wavefunction for a single exciton, 
\begin{equation}
\Big[\frac{\hbar^2k^2}{2m}\delta_{\vec{k}\vec{k}'}
-\frac{1}{A}U(\vec{k}-\vec{k}')\Big]f(\vec{k}')=-E_b f(\vec{k}),
\label{Xwave}
\end{equation}
where the reduced mass $m=m_e m_h/M$, and $E_b$ is the binding energy for $1s$ state.
The normalization condition is that
\begin{equation}
\frac{1}{A}\sum_{\vec{k}}f(\vec{k})^2=1.
\end{equation}
Here we have chosen $f(\vec{k})$ to be real.
In Eq.~(\ref{LB}), $f(\vec{k}+x_h \vec{Q})$ is the wavefunction for an exciton with center-of-mass momentum $\vec{Q}$.

By substituting Eq.~(\ref{LB}) into Eq.~(\ref{BPhase}) and (\ref{En4}), we obtain that
\begin{equation}
\mathcal{B}\approx\frac{i}{2}\big(B_{(vc)\vec{Q}}^*\partial_t B_{(vc)\vec{Q}}-B_{(vc)\vec{Q}}\partial_t B_{(vc)\vec{Q}}^*\big),
\label{BBB}
\end{equation}
\begin{equation}
\begin{aligned}
&\mathcal{H}=\sum\big(\frac{\hbar^2Q^2}{2M}-E_b-\mu\big)B_{(v c)\vec{Q}}^* B_{(v c)\vec{Q}}\\
+&\frac{1}{2A}\sum'\big\{ g_H(\vec{Q}_{14})B_{(v c)\vec{Q}_1}^* B_{(v' c')\vec{Q}_2}^* B_{(v' c')\vec{Q}_3}B_{(v c)\vec{Q}_4}\\
+&g_X(\vec{Q}_{13},\vec{Q}_{14})B_{(v c)\vec{Q}_1}^* B_{(v' c')\vec{Q}_2}^* B_{(v' c)\vec{Q}_3}B_{(v c')\vec{Q}_4}\big\}.
\end{aligned}
\label{HBoson}
\end{equation}
The Berry phase $\mathcal{B}$  has the
same form as that in the field-theory functional integral
representation of a standard interacting boson model, which suggests that the Lagrangian $\mathcal{L}=\mathcal{B}-\mathcal{H}$ is a functional field integral representation of a boson model\cite{Negele1988}.
By replacing the complex numbers $(B^*, B)$ with bosonic creation and annihilation operators $(B^\dagger, B)$ in the energy functional $\mathcal{H}$,
we arrive at the boson model (\ref{HBoson0}) in the main text.

$g_H(\vec{q})$ is derived from $\mathcal{H}_H^{(4)}$, and has the following analytic expression
\begin{equation}
\begin{aligned}
g_H(\vec{q})&=V(\vec{q})\big[F(x_e \vec{q})^2+F(x_h \vec{q})^2\big]\\
&-2U(\vec{q})F(x_e \vec{q})F(x_h \vec{q}),
\end{aligned}
\label{gH}
\end{equation}
where
\begin{equation}
F(\vec{q})=\frac{1}{A}\sum_{\vec{k}}f(\vec{k})f(\vec{k}+\vec{q}).
\end{equation}
For zero-momentum transfer, $g_H(0)=2(V(0)-U(0))=4\pi e^2 d/\epsilon$.

\begin{figure}[t]
	\includegraphics[width=0.7\columnwidth]{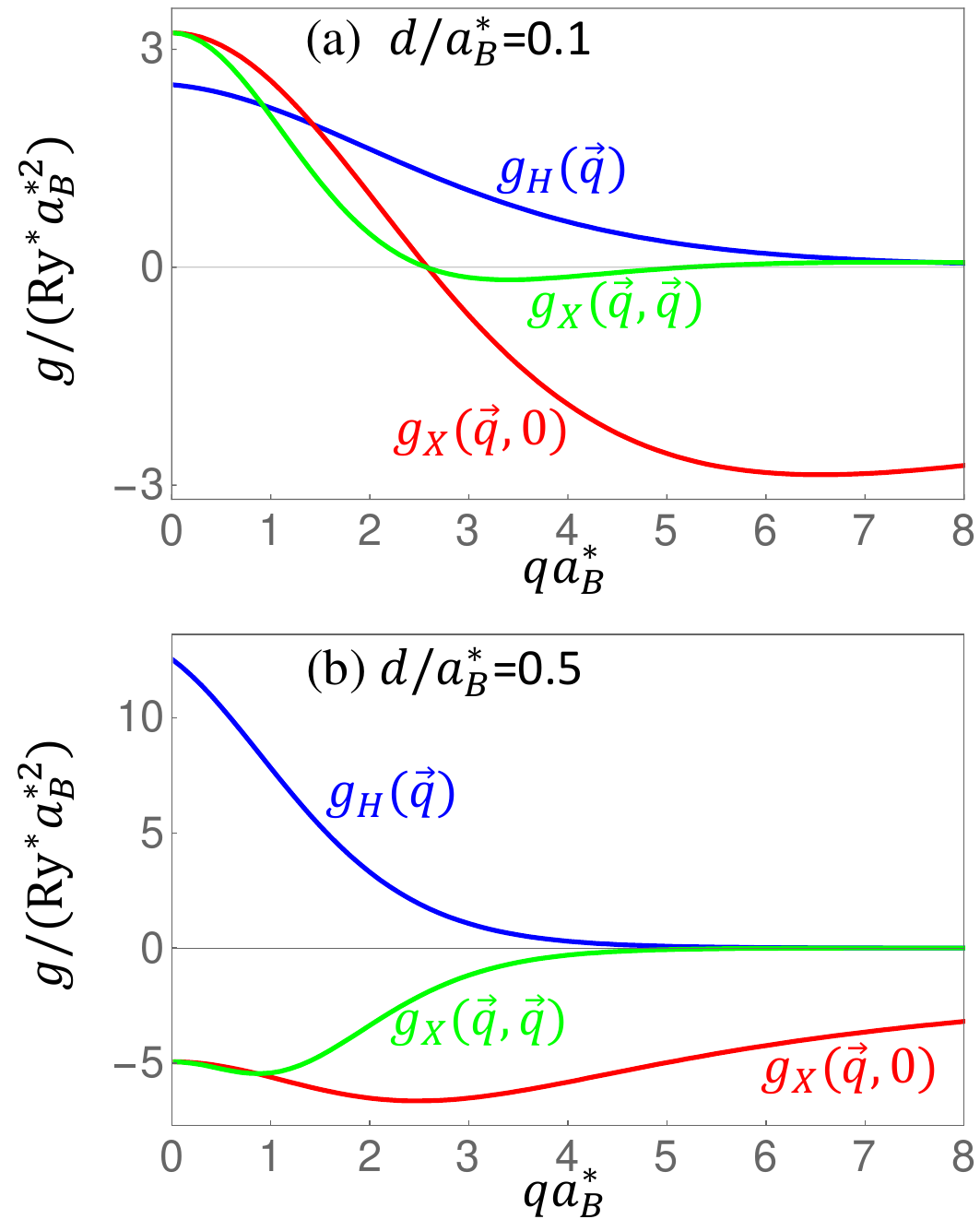}
	\caption{ $g_H(\vec{q})$, $g_X(\vec{q}, 0)$ and $g_X(\vec{q}, \vec{q})$ as a function of momentum. $d/a_B^*$ is 0.1 in (a) and 0.5 in (b). $m_e=m_h$ is assumed.
	}
	\label{Fig:XII}
\end{figure}

$g_X(\vec{Q}_{13},\vec{Q}_{14})$ is derived from $\mathcal{H}_X^{(4)}$, and can be further decomposed into two parts $g_X=g_X^{(1)}-g_X^{(2)}$.
$g_X^{(1)}$ arises from the loss of interlayer exchange energy as more excitons condense, while $-g_X^{(2)}$ comes from the gain of intralayer exchange energy as electron or hole density increases. The explicit forms of $g_X^{(1,2)}$ are below.
\begin{equation}
\begin{aligned}
&g_X^{(1)}(\vec{Q}_{13},\vec{Q}_{14})\\
=&\frac{1}{A^2}\sum_{\vec{k},\vec{k}'}U(\vec{k}-\vec{k}')f(\vec{k}')f(\vec{k}+\vec{Q}_e)\\
&\times\big[f(\vec{k}+\vec{Q}_h)f(\vec{k}+\vec{Q}_e+\vec{Q}_h)\\
&+f(\vec{k}-\vec{Q}_h)f(\vec{k}+\vec{Q}_e-\vec{Q}_h)\big]\\
=&\frac{1}{A}\sum_{\vec{k}}\big(E_b+\frac{\hbar^2k^2}{2m}\big)f(\vec{k})f(\vec{k}+\vec{Q}_e)\\
&\times\big[f(\vec{k}+\vec{Q}_h)f(\vec{k}+\vec{Q}_e+\vec{Q}_h)\\
&+f(\vec{k}-\vec{Q}_h)f(\vec{k}+\vec{Q}_e-\vec{Q}_h)\big],
\end{aligned}
\label{gX1}
\end{equation}
\begin{equation}
\begin{aligned}
&g_X^{(2)}(\vec{Q}_{13},\vec{Q}_{14})
=\frac{1}{A^2}\sum_{\vec{k},\vec{k}'}\big\{\\
&V(\vec{k}-\vec{k}'+\vec{Q}_e)f(\vec{k})f(\vec{k}')f(\vec{k}+\vec{Q}_h)f(\vec{k}'+\vec{Q}_h)+\\
&V(\vec{k}-\vec{k}'+\vec{Q}_h)f(\vec{k})f(\vec{k}')f(\vec{k}+\vec{Q}_e)f(\vec{k}'+\vec{Q}_e)
\big\},
\end{aligned}
\label{gX2}
\end{equation}
where $\vec{Q}_e=x_e\vec{Q}_{13}$ and $\vec{Q}_h=x_h\vec{Q}_{14}$.
According to Eq.~(\ref{gX1}), $g_X^{(1)}$ can also be interpreted as the increase of kinetic energy due to the increase of the exciton density.
The diagrammatic representation of different processes for the exciton-exciton interaction is shown in 
Fig.~\ref{Fig:HI} and \ref{Fig:XI}. The momentum dependence of the interaction strength $g_{H, X}$ is illustrated in Fig.~\ref{Fig:XII}.
Both $g_{H}$ and $g_{X}$ have a strong momentum dependence, which indicates that extion-exciton interaction are long-ranged instead of short-ranged.

The problem of exciton-exction interaction has been studied using many different methods in the literature\cite{Hanamura1970,  Laikhtman2001, Okumura2001, Zimmermann2008, Dubin2008, Dubin2015}. Here  we used an alternative approach based on a variational wave function combined with the Lagrangian formalism. The functional field integral representation of the boson model in Eq.~(\ref{HBoson0}) is exactly given by the Lagrangian $\mathcal{L}$ with Berry phase $\mathcal{B}$ in Eq.~(\ref{BBB}) and energy function $\mathcal{H}$ in Eq.~(\ref{HBoson})\cite{Negele1988}. This property establishes the equivalence between the Lagrangian and the boson model. In fact, all results obtained using the boson model can be equivalently derived from the Lagrangian. The phase transition between phase-I and II can be determined by the minimization of the energy functional (\ref{HBoson}) with the ansatz that $(B^*, B)$ is spatially uniform and time independent.  The collective mode studied in Appendix \ref{CModeBoson} can be obtained using the Euler-Lagrange equation of the Lagrangian.

There are other approaches in constructing an effective theory of bosonic excitation in an interacting fermion system. One example is the 
standard Hubbard-Stratonovich transformation. There is however a certain arbitrariness in decomposing electron-electron interactions into Hartree or Fock channels in the Hubbard-Stratonovich scheme \cite{Negele1988}. The HS approach is not appropriate here because a proper description of the SIEXC requires that Hartree and Fock interactions to be treated on the same footing.

\section{Interacting boson model for excitons with zero center-of-mass momentum}
\label{App2}
We consider another variational wave function with all excitons condense into zero center-of-mass momentum state.
\begin{equation}
\begin{aligned}
&|\Psi_0\rangle=\prod_{\vec{k}}\big[s_1(\vec{k})a_{v_1\vec{k}}^\dagger
+f(\vec{k})\sum_{c}b_{(v_1c)}a_{c\vec{k}}^\dagger\big]\times\\
&\big[\chi(\vec{k})a_{v_1\vec{k}}^\dagger+s_2(\vec{k})a_{v_2\vec{k}}^\dagger
+f(\vec{k})\sum_{c}b_{(v_2c)}a_{c\vec{k}}^\dagger\big]|0\rangle,
\end{aligned}
\end{equation}
where $v_1=1$ and $v_2=2$, representing two valence bands, and $c$ represents different conduction bands. $f(\vec{k})$ is the exciton wave function in Eq.~(\ref{Xwave}). $b_{(vc)}$ are complex parameters, which are independent of momentum $\vec{k}$. In $|0\rangle$, both valence and conduction bands are empty, $a_{c\vec{k}}|0\rangle=a_{v\vec{k}}|0\rangle=0$.
Facotrs $s_{1,2}(\vec{k})$ and $\chi(\vec{k})$ are determined by normalization and orthogonality conditions. 
By normalization conditions, we have that
\begin{equation}
\begin{aligned}
s_1(\vec{k})&=\sqrt{1-f(\vec{k})^2\sum_{c}|b_{(v_1c)}|^2},\\
s_2(\vec{k})&=\sqrt{1-|\chi(\vec{k})|^2-f(\vec{k})^2\sum_{c}|b_{(v_2c)}|^2}.
\end{aligned}
\end{equation}
To ensure that at each momentum $\vec{k}$ the two occupied states are orthogonal to each other, we require that 
\begin{equation}
s_1(\vec{k})\chi(\vec{k})+f(\vec{k})^2\sum_{c}b_{(v_1c)}^*b_{(v_2c)}=0.
\end{equation}

Similar to the procedure in App.~\ref{sec.BModel}, we expand the energy functional $\langle\Psi_0 | H | \Psi_0 \rangle$ to fourth order in $b_{(vc)}$, and then replace complex numbers $(b_{(vc)}^*, b_{(vc)})$ by operators $(B_{(vc)\vec{0}}^\dagger, B_{(vc)\vec{0}})/\sqrt{A}$, which gives rise to the same boson model in Eq.~(\ref{HBoson}) except that all momenta are restricted to be zero.

\section{Collective modes in the interacting boson model}
\label{CModeBoson}
The collective modes can be calculated analytically using the interacting boson model. The strategy is to shift a bosonic operator by its mean-field value,
\begin{equation}
B_{(vc)\vec{Q}}=\langle B_{(vc)\vec{Q}} \rangle \delta_{\vec{Q}, 0}+b_{(vc)\vec{Q}},
\end{equation}
where $b_{(vc)\vec{Q}}$ is also a bosonic operator that describes fluctuations around the mean field state.
The bosonic Hamiltonian in Eq.~(\ref{HBoson0}) is then expanded to second order in $(b^\dagger, b)$.
The first order terms vanish as the energy functional in Eq.~(\ref{eq:totalenergy}) is minimized in the mean-field state. 
The quadratic terms can be diagonalized using the Bogoliubov transformation, giving rise to the collective mode spectra. 
We present the dispersion and degeneracy of collective modes in phase-I and II without giving the details of the derivation.

In phase-I, there are five gapless modes, and three gapped modes. Given the mean-field value in Eq.~(\ref{FI}), $(b^\dagger, b)_{(vc)}$ with different composite index $(vc)$ are decoupled. The $(vc)=(11)$ mode is gapless, which is the Goldstone mode due to the spontaneously broken U(1) symmetry, i.e. the separate charge conservation within each layer. Its dispersion is:
\begin{equation}
\begin{aligned}
&\omega_{(11)}=\sqrt{\Big[\frac{\hbar^2Q^2}{2M}+(2g_H+g_+)(\vec{Q})n_\text{I}\Big]\Big[\frac{\hbar^2Q^2}{2M}+g_-(\vec{Q})n_\text{I}\Big]},\\
&g_{\pm}(\vec{Q})=g_X(\vec{Q},0)+g_X(0,\vec{Q})-g_X(0,0)\pm g_X(\vec{Q},\vec{Q}),
\end{aligned}
\end{equation}
where $g_{-}(\vec{Q})$ has a $Q^2$ dependence at small $\vec{Q}$. Therefore, the (11) mode has a linear dispersion in $Q \rightarrow 0$ limit.

The $(vc)=(12)$, (13) and (14) modes are degenerate, with a gapless dispersion:
\begin{equation}
\omega_{(12)}=\frac{\hbar^2Q^2}{2M}+\big[ g_x(0, \vec{Q}) - g_x(0, 0) \big]n_\text{I}, 
\end{equation}
which has a quadratic $Q$ dependence in $Q \rightarrow 0$ limit.
Similaryly, the $(vc)=(21)$ mode is also gapless and quadratic at small $\vec{Q}$:
\begin{equation}
\omega_{(21)}=\frac{\hbar^2Q^2}{2M}+\big[ g_x( \vec{Q}, 0) - g_x(0, 0) \big]n_\text{I}.
\end{equation}

The $(vc)=(22)$, (23) and (24) modes are degenerate and gapped:
\begin{equation}
\omega_{(22)}=\frac{\hbar^2Q^2}{2M} - g_x(0, 0) n_\text{I}, 
\end{equation}
which shows that phase-I is stable against small fluctuations provided that $g_x(0, 0)$ is negative.

In phase-II, the mean-field values are given in Eq.~(\ref{FII}) and all eight collective modes are gapless. The $(vc)=$ (11) and (22) fluctuations are coupled, and give rise to two non-degenerate gappless modes:
\begin{equation}
\begin{aligned}
\omega_{(11), (22)}&=\sqrt{\Big[\frac{\hbar^2Q^2}{2M}+(4g_H+g_+)(\vec{Q})\frac{n_\text{II}}{2}\Big]}\\
&\times\sqrt{\Big[\frac{\hbar^2Q^2}{2M}+g_-(\vec{Q})\frac{n_\text{II}}{2}\Big]},\\
\omega_{(11), (22)}'&=\sqrt{\Big[\frac{\hbar^2Q^2}{2M}+g_+(\vec{Q})\frac{n_\text{II}}{2}\Big]}\\
&\times\sqrt{\Big[\frac{\hbar^2Q^2}{2M}+g_-(\vec{Q})\frac{n_\text{II}}{2}\Big]},\\
\end{aligned}
\end{equation}
both of which have a linear dispersion at small $Q$.

The $(vc)=$ (12) and (21) fluctuations are also coupled, and lead to two modes that are degenerate with $\omega_{(11), (22)}'$.

The $(vc)=$ (13), (14), (23) and (24) fluctuations are decoupled, and have degenerate collective modes:
\begin{equation}
\omega_{(13)}=\frac{\hbar^2Q^2}{2M}+\big[ g_x(0, \vec{Q}) - g_x(0, 0) \big]\frac{n_\text{II}}{2},
\end{equation}
which has a quadratic dispersion at small $Q$.

\begin{table}[hbt]
	\caption{Classification of collective modes in phase-I and II. $N_{\text{gapped}}$ is the number of gapped collective modes.  $N_1$ and  $N_2$ are respectively the number of gapless collective modes with linear and quadratic dispersion. $N_{\text{BSG}}$ is the number of the broken symmetry generators.}
		\begin{tabular}{ l | c | c | c | c }
			\hline
			& $N_{\text{gapped}}$ & $N_1$ & $N_2$ & $N_{\text{BSG}}$\\
			\hline
			phase-I & 3  & 1 & 4 &  9  \\
			\hline
			phase-II & 0 & 4 & 4 &  12  \\
			\hline
	\end{tabular}		
	\label{Classification}
\end{table}

In Table ~\ref{Classification}, we list $N_1$ and  $N_2$, respectively the number of gapless collective modes with linear and quadratic dispersion, and $N_{\text{BSG}}$, the number of the broken symmetry generators for each phase. There is a relationship among these three numbers in both phase-I and II,
\begin{equation}
N_1+2N_2=N_{\text{BSG}}.
\end{equation} 
This relationship is typical for broken symmetry states in system without Lorentz invariance\cite{Watanabe2012}.

\begin{widetext}
\section{explicit expressions for $\mathcal{E}_{\vec{k},\vec{p}}(\vec{Q})$ and $\Gamma_{\vec{k},\vec{p}}(\vec{Q})$}
\label{App3}
Below are explicit expressions for $\mathcal{E}_{\vec{k},\vec{p}}(\vec{Q})$ and $\Gamma_{\vec{k},\vec{p}}(\vec{Q})$, which appear in the energy variation $\delta E^{(2)}$ in Eq.~(\ref{EF}).
\begin{equation}
\begin{split}
\mathcal{E}_{\vec{k},\vec{p}}(\vec{Q})&=\delta_{\vec{k},\vec{p}}(\zeta_{\vec{k}+\vec{Q}}-\zeta_{\vec{k}}+E_{\vec{k}}+E_{\vec{k}+\vec{Q}})\\
&+\frac{1}{A}\big[V(\vec{Q})-V(\vec{k}-\vec{p})\big]
(u_{\vec{k}}u_{\vec{p}}v_{\vec{k}+\vec{Q}}v_{\vec{p}+\vec{Q}}
+v_{\vec{k}}v_{\vec{p}}u_{\vec{k}+\vec{Q}}u_{\vec{p}+\vec{Q}})\\
&-\frac{1}{A}U(\vec{Q})(v_{\vec{k}}u_{\vec{p}}u_{\vec{k}+\vec{Q}}v_{\vec{p}+\vec{Q}}+u_{\vec{k}}v_{\vec{p}}v_{\vec{k}+\vec{Q}}u_{\vec{p}+\vec{Q}})\\
&-\frac{1}{A}U(\vec{k}-\vec{p})(u_{\vec{k}}u_{\vec{p}}u_{\vec{k}+\vec{Q}}u_{\vec{p}+\vec{Q}}+v_{\vec{k}}v_{\vec{p}}v_{\vec{k}+\vec{Q}}v_{\vec{p}+\vec{Q}}),\\
\Gamma_{\vec{k},\vec{p}}(\vec{Q})&=\frac{1}{A}\big[V(\vec{Q})-V(\vec{k}+\vec{Q}-\vec{p})\big]
(u_{\vec{k}}u_{\vec{p}}v_{\vec{k}+\vec{Q}}v_{\vec{p}-\vec{Q}}
+v_{\vec{k}}v_{\vec{p}}u_{\vec{k}+\vec{Q}}u_{\vec{p}-\vec{Q}})\\
&-\frac{1}{A}U(\vec{Q})(v_{\vec{k}}u_{\vec{p}}u_{\vec{k}+\vec{Q}}v_{\vec{p}-\vec{Q}}
+u_{\vec{k}}v_{\vec{p}}v_{\vec{k}+\vec{Q}}u_{\vec{p}-\vec{Q}})\\
&+\frac{1}{A}U(\vec{k}+\vec{Q}-\vec{p})
(v_{\vec{k}}u_{\vec{p}}v_{\vec{k}+\vec{Q}}u_{\vec{p}-\vec{Q}}
+u_{\vec{k}}v_{\vec{p}}u_{\vec{k}+\vec{Q}}v_{\vec{p}-\vec{Q}}),
\end{split}
\end{equation}
where $u_{\vec{k}}$ and $v_{\vec{k}}$ are defined in Eq.~(\ref{uv}), and $V(\vec{Q})$ and $U(\vec{Q})$ are respectively intralayer and interlayer Coulomb interactions.
\end{widetext}

\bibliography{SIEXC}{}
\bibliographystyle{apsrev4-1}

\end{document}